\documentclass[journal]{IEEEtran}
\ifCLASSINFOpdf
\usepackage[pdftex]{graphicx}
\else
\fi
\usepackage{amsmath}
\usepackage{color}
\usepackage{multirow}
\usepackage{array}
\usepackage{algpseudocode}
\usepackage{algorithm}
\usepackage{algorithmicx}
\usepackage{algpseudocode}
\usepackage{amsmath}
\usepackage{listings}

\usepackage{amsmath}
\usepackage{amssymb}
\usepackage{amsthm}
\usepackage{graphicx}
\usepackage{algorithm}
\usepackage{algpseudocode}

\begin{document}

%
\title{An Efficient NPN Boolean Matching Algorithm Based on Structural Signature and Shannon Expansion}
%
%
%

\author{Juling Zhang,
        Guowu Yang,
        William N. N. Hung,
        Yan Zhang,
        and Jinzhao Wu         
\thanks{J.L. Zhang, G.W. Yang and Y. Zhang are with Big Data Research Center, School of Computer Science and Engineering, University of Electronic Science and Technology of China, Chengdu, 611731, China. E-mail: (zjlgj@163.com, guowu@uestc.edu.cn, yixianqianzy@gmail.com).}
\thanks{W.N.N. Hung is with Synopsys Inc., Mountain View, CA, USA. E-mail:( William.Hung@synopsys.com).}
\thanks{J.Z. Wu is with Guangxi Key Laboratory of Hybrid Computation and IC Design Analysis, Guangxi University for
Nationalities, Nanning, 530006, China. E-mai: (gxmdwjzh@aliyun.com)}}

%
%

\markboth{Journal of \LaTeX\ Class Files,~Vol.~14, No.~8,  November~2016}%
{Shell \MakeLowercase{\textit{et al.}}: Bare Demo of IEEEtran.cls for IEEE Journals}
%



\maketitle

\begin{abstract}
An efficient pairwise Boolean matching algorithm for solving the problem of matching single-output specified Boolean functions under input negation and/or input permutation and/or output negation (NPN) is proposed in this paper. We present the structural signature (SS) vector, which comprises a first-order signature value, two symmetry marks, and a group mark. As a necessary condition for NPN Boolean matching, the SS is more effective than the traditional signature. Two Boolean functions, $f$ and $g$, may be equivalent when they have the same SS vector. A symmetry mark can distinguish symmetric variables and asymmetric variables and be used to search for multiple variable mappings in a single variable-mapping search operation, which reduces the search space significantly. Updating the SS vector via Shannon decomposition provides benefits in distinguishing unidentified variables, and the group mark and phase collision check can be used to discover incorrect variable mappings quickly, which also speeds up the NPN Boolean matching process. Using the algorithm proposed in this paper, we test both equivalent and non-equivalent matching speeds on the MCNC benchmark circuit sets and random circuit sets. In the experiment, our algorithm is shown to be 4.2 times faster than competitors when testing equivalent circuits and 172 times faster, on average, when testing non-equivalent circuits. The experimental results show that our approach is highly effective at solving the NPN Boolean matching problem.
\end{abstract}

\begin{IEEEkeywords}
Boolean Matching, Structral Signature Vector, Shannon Expansion, Variable Mapping, Variable Symmetry.
\end{IEEEkeywords}

%
\IEEEpeerreviewmaketitle

\section{Introduction}
%
%
%
%
NPN Boolean matching is a significant problem in logic synthesis, technology mapping, cell-library binding, and logic verification \cite{15}. In logic verification, a key issue is to verify the equivalence of two circuit functions~\cite{8}. In both cell-library binding and technology mapping, a Boolean network is transformed into an equivalent circuit using logic cells from a standard cell library~\cite{14,10}. Boolean matching is a critical step in technology mapping, as demonstrated in \cite{17}. NPN Boolean matching can be applied to either incompletely or completely specified functions with either multiple or single outputs. In this paper, we study the NPN Boolean matching problem for completely specified functions with a single output.

The NPN Boolean matching problem involves judging whether one Boolean function can be transformed into another via input negation and/or input permutation and/or output negation. If the Boolean function $g$ can be obtained from $f$ via one of the previous types of transformations, then $f$ and $g$ are NPN-equivalent; that is, the logic circuit of $g$ could be implemented by $f$.

An $n$-variable Boolean function has ${2^n}$ input negations, $n!$ input permutations, and 2 possible output negations. The complexity of the exhaustive method for NPN Boolean matching is $O(n!{2^{n + 1}})$. However, a Boolean function with numerous input variables is intractable using the exhaustive approach. In this paper, an effective pairwise NPN Boolean matching algorithm is proposed. A binary decision diagram (BDD) is used to represent a Boolean function because this representation is compact and previous studies have demonstrated its high speed when computing signatures~\cite{14}. The algorithm proposed in this paper utilizes a structural signature (SS) vector and Shannon expansion to detect possible transformations of two Boolean functions. In our proposed Boolean function matching algorithm, the SS vector is required. An SS value comprises a first-order signature value, two symmetry marks, and a group mark. When the SS values of any two variables are the same, variable mappings between the two variables may exist. The symmetry marks serve to distinguish symmetric and asymmetric variables. They allow the removal of many impossible variable mappings and can be used to find multiple variable mappings in one variable-mapping search operation, which reduces the search space of Boolean matching. A group mark can be used to solve the first-order signature value collision problem.
Boolean decomposition updates the SS vectors. The updated SS vectors are then used to further search for variable mappings. Phase collision check can be carried out to find error variable mappings quickly.
All these methods are used to speed up transformation detection. For two NPN-equivalent Boolean functions $f$ and $g$, the goal of our algorithm is to find a transformation that can transform $f$ into $g(\overline{g})$ as quickly as possible. In the algorithm, a tree structure is used to store the detected transformations. It addresses transformations in depth-first search (DFS) order.

In cell-library binding, a logic cell can be found in a standard cell library that can implement an objective circuit. During cell-library binding, the probability of having the same SS vectors is lower, and our algorithm operates very quickly when judging non-equivalent Boolean functions. Therefore, our algorithm can be applied to cell-library binding. The same is true when applying our algorithm to technology mapping.

A BDD is used in both the algorithm proposed in \cite{3} and our algorithm to represent a Boolean function and to compute the cofactor and symmetry. Because the results of \cite{3} reflect the state of the art in NPN Boolean matching using BDD, we reimplement the algorithm from \cite{3} and compare its runtime with that of our algorithm on the same data set. We test the algorithm on both MCNC benchmark circuits and random circuits and report the runtimes for equivalent and non-equivalent Boolean matching. By comparing the results of our algorithm with those of the algorithm from \cite{3}, we conclude that our proposed algorithm runs twice as fast as the algorithm from \cite{3} for general equivalent circuits, while for non-equivalent circuits, our algorithm is much faster than the algorithm proposed in \cite{3}. Therefore, the proposed algorithm is highly effective. Moreover, it can scale to at most 22 variables in NPN Boolean matching.

The paper is organized as follows. Section \uppercase\expandafter{\romannumeral2} presents current research results concerning NPN Boolean matching. Section \uppercase\expandafter{\romannumeral3} introduces related preliminaries and defines the problem to be solved. The NPN Boolean matching algorithm is explained in Section \uppercase\expandafter{\romannumeral4}. In Section \uppercase\expandafter{\romannumeral5}, we provide experimental results that verify the effectiveness of our algorithm. Finally, we summarize our work and outline directions for future research in Section \uppercase\expandafter{\romannumeral6}.

\section{Related Works}
Scholars have explored numerous methods and algorithms for solving the Boolean matching problem. Boolean matching algorithms are traditionally categorized into two classes: (1) algorithms based on canonical forms and (2) algorithms based on pairwise matching~\cite{3}. The key to the first approach is to construct a unified canonical form for equivalent Boolean functions. Two Boolean functions are equivalent when their canonical forms are identical. The second method is a semi-exhaustive search in which some appropriate signatures are used to establish variable mapping and reduce the search space. Function signatures are properties of Boolean functions that are independent of input variable permutation and negation~\cite{6}, and they are intrinsic characteristics of Boolean functions~\cite{11}.

In Boolean matching based on canonical forms, a unique Boolean function is used as a representative for each Boolean equivalent class. References \cite{4}, \cite{5}, and \cite{20} introduced NPN Boolean matching based on canonical forms. Their canonical representative was a Boolean function with a maximal truth table. Variable symmetry and DFS were utilized to reduce the search complexity in~\cite{4}. Using a look-up table and a tree-based breadth-first search, the authors of ~\cite{20} presented an algorithm that can be used to compute an NP-representative. Ciric and Sechen ~\cite{2} proposed an efficient canonical form for P-equivalent Boolean matching. The canonical form of~\cite{2} involves the lowest cost required to find the representative under a permutation of rows and columns in a truth table. Damiani and Selchenko~\cite{17} used decomposition trees to create the Boolean function representative. Agosta et al.~\cite{15} combined the spectral and canonical forms to exploit a transform-parametric approach that could match Boolean functions of up to 20 variables, but this method considers only the input permutation. The authors of \cite{27} proposed a fast Boolean function NPN classification using canonical forms, in which a Boolean function is represented by a truth table. The method then computes the canonical form using a cofactor, swapping variables and symmetry. Petkovska et al.~\cite{28} proposed a hierarchical method to resolve NPN Boolean classification that improved classification speed compared with ~\cite{27}. However, these two papers tested NPN classification only for functions with 6--16 inputs. Abdollahi and Pedram~\cite{16} proposed a high-order signature and presented a P-equivalent Boolean matching algorithm. Agosta et al.~\cite{9} used shifted cofactor transformation to achieve more efficient P-equivalent matching. Abdollahi~\cite{3} constructed a NPN Boolean matching algorithm utilizing the signature-based canonical form and extended it to incompletely specified Boolean matching~\cite{14}. Although the experimental result in \cite{3} was achieved quickly, only the execution time needed to establish the canonical form was published. Therefore, in this study, we reimplemented the algorithm from \cite{3} and compared it to the runtime of our NPN Boolean matching algorithm.

The key to creating a pairwise matching algorithm is to establish a one-to-one variable correspondence between two Boolean functions. In this method, the cofactor signature is universally used to search for variable correspondence. However, some variable correspondences cannot be determined using only the cofactor signature. Mohnke and Malik~\cite{6} exploited an NPN Boolean matching algorithm that uses a breakup signature and Boolean difference when the cofactor signature fails. The authors of ~\cite{3} expanded the cofactor signature to an $n$-order signature. Lai et al.~\cite{1} used a level-first search to complete Boolean matching and proposed a set of filters that can be used to improve the algorithm performance. The consensus and smoothing operators were applied to Boolean matching with a don't care set by Chen~\cite{8}. Wang and Hwang~\cite{21} developed an efficient Boolean matching algorithm that uses cofactor and equivalence signatures based on the communication complexity. The authors of \cite{10,12,13,14} presented a pairwise Boolean matching that utilizes don't care sets in technology mapping.

Other approaches to Boolean matching exist. SAT-based Boolean matching has been used for the technology mapping of specific circuits; see, e.g.,~\cite{7}, \cite{22}, and \cite{23}. Yu et al.~\cite{7} developed a filter-based Boolean matcher with FC-BM and FH-BM that can solve the NPN-equivalence in FPGA. However, this approach can only handle 9 variables. Based on functional decomposition operations, Cong and Hwang~\cite{18} exploited a Boolean matching method for LUT-based PLBs. Soeken et al. \cite{29} proposed a heuristic NPN classification approach for large functions that involves AIGs and LEXSAT. The authors of \cite{30} researched graph-based, simulation-driven, and SAT-based PP Boolean matching, implementing PP Boolean equivalence-checking for large-scale circuits. Many Boolean matching studies have utilized symmetry because it can effectively reduce the search space. The authors of \cite{31} proposed a generalized Boolean symmetry and applied it to PP Boolean matching for large circuits. Lai et al.~\cite{24} proposed Boolean matching with strengthened learning.

\section{Preliminaries and Problem Statement}

In this paper, we use $\emph{X}=(x_0,x_1,...x_{n-1})$ to express a vector of Boolean variables and $f(X):B^n \to B$ to express a single-output, completely specified Boolean function. Some of the following definitions were also presented in \cite{3}.

\emph{Definition 1 (Cofactor of Boolean Function):} The cofactor of Boolean function $f$ with respect to variable ${x{}_i({{\overline {x}}_i}) }$ is expressed as ${{f_{{x_i}}}({f_{{{\overline {x}}_i} }})}$, which is computed by substituting ${x{}_i({{\overline {x}}_i}) }=1$ into $f$.

${{f_{{x_i}}}({f_{{{\overline {x}}_i} }})}$ is a new Boolean function of $n-1$ variables. Let ${\left| {{f_{{x_i}}}} \right|}$ denote the number of minterms covered by ${{f_{{x_i}}}}$ and $\left| {{f_{{{\overline {x}}_i} }}} \right|$ denote the number of minterms covered by ${f_{{{\overline {x}}_i} }}$.

A cube b is the conjunction of some variables, which can be positive or negative. For any cube $b$, ${f_b}$ is the cofactor of $f$ with respect to $b$, and $\left| {f_b} \right|$ denotes the number of minterms covered by ${f_b}$.

\emph{Lemma 1: }Shannon expansion is the identity ${f = {x_i}{f_{{x_i}}} + {{\overline {x}}_i} {f_{{{\overline {x}}_i} }}}$ (also referred to as Shannon decomposition).

The equivalence of two Boolean functions can be verified via recursive Shannon decomposition.

\emph{Definition 2 (NP Transformation):} An NP transformation $T$ is composed of the negation and/or permutation of input variables. Consider a Boolean vector  $X=(x_0,x_1,...x_{n-1})$. Its NP transformation can be expressed as $T = \left( {\begin{array}{*{20}{c}}
{{x_0},{x_1},...,{x_{n - 1}}}\\
{x_{\sigma (0)}^{{\alpha _0}},x_{\sigma (1)}^{{\alpha _1}},...,x_{\sigma (n - 1)}^{{\alpha _{n - 1}}}}
\end{array}} \right),{\alpha _i} \in \{ 0,1\} $, where $\sigma$ is a permutation of $(0,1,...,n-1)$ and $\alpha _i$ indicates whether $x_i$ takes a negation operation, such as $x_i^1 = {x_i},x_i^0 = {{\overline {x}}_i} $.

\emph{Example 1:} Consider a Boolean function $f({x_0},{x_1},{x_2}) = {{\overline{x}}_0}{{\overline{x}}_1}{x_2} + {x_0}{{\overline{x}}_1}{{\overline {x}}_2}$ and a transformation $T = \left( {\begin{array}{*{20}{c}}
{{x_0},{x_1},{x_2}}\\
{x_2^0,x_1^0,x_0^0}
\end{array}} \right)$. If $g({x_0},{x_1},{x_2}) = f(TX) = {{\overline {x}}_0} {x_1}{x_2} + {x_0}{x_1}{{\overline {x}}_2}$, then $f$ is NP-equivalent to $g$.

\emph{Definition 3 (NPN-Equivalence):} Two Boolean functions $f(X)$ and $g(X)$ are NPN-equivalent if there exists an NP transformation $T$ that satisfies $f(TX)=g(X)$ or $f(TX)=\overline{g(X)}$.

\emph{Definition 4 (Zeroth-order Signature Value):} The zeroth-order signature value of function $f$ is the number of its minterms. In this paper, the zeroth-order signature value is expressed as $\left| f \right|$ for $f(X)$.

\emph{Definition 5 (First-order Signature Value):} The first-order ($1^{st}$) signature value of function $f$ with respect to ${x{}_i}$ is ${V^i}=(\left| {{f_{{x_i}}}} \right|,\left| {{f_{{{\overline {x}}_i} }}} \right|)$. The first-order signature vector of $f$ is denoted as ${V^f} = \{ { ( \left| {{f_{{x_i}}}} \right|,\left| {{f_{{{\overline {x}}_i} }}} \right| )}|i \in \ \{0,1, \cdots, n - 1\}\} $.

Because the signature is the key characteristic of Boolean functions, having the same signature is often regarded as a necessary condition for the NP-equivalence of Boolean functions~\cite{14}. If two Boolean functions are NP-equivalent, then they must have the same $1^{st}$ signature vector (without consideration of variable order), which is also a necessary condition for NP Boolean equivalence prediction.

\emph{Example 2:} Consider the Boolean functions $f$, $g$ and $h$:

${f(X) = {x_0}{{\overline{x}}_1}+ {{\overline {x}}_1}{x_2} + {{\overline {x}}_0}{x_1}{{\overline {x}}_2}}$

${g(X) = {x_0}{x_2} + {x_1}{x_2} + {{\overline {x}}_0}{{\overline {x}}_1}{{\overline {x}}_2}}$

${h(X) = {x_0}{{\overline {x}}_1}  + {x_0}{x_2} + {{\overline {x}}_1} {x_2}}$

The $1^{st}$ signature vectors of the three Boolean functions are, respectively, as follows: ${V^f}$=${\{(2,2),(1,3),(2,2)\}}$, ${V^g}$=${\{(3,1),(2,2),(2,2)\}}$, and ${V^h}$=${\{(3,1),(1,3),(3,1)\}}$. Because having the same $1^{st}$ signature vector is a necessary condition for NP Boolean equivalence, it can be concluded that $f$ and $h$ must not be NP-equivalent, while Boolean functions $f$ and $g$ may be NP-equivalent.

Consider two NP-equivalent Boolean functions $f$ and $g$ and the NP transformation $T$ between them. The $1^{st}$ signature value of variable $x_i$ of $f$ must be same as the $1^{st}$ signature value of variable $x_j$ of $g$ if $x_i$ is permuted to $x_j({\overline{x}}_j)$ of $T$.

\emph{Definition 6 (Variable Mapping):} Variable mapping is the correspondence relation between two variables of two Boolean functions. It is a permutation of variable ${x_i}$ and ${x_j}({{\overline {x}}_j)} $. The variable mapping of a specified variable, ${x_i}$, is ${\varphi _i}:{x_i} \to x_j^{{\alpha}}$, where $i,j \in \{0,1, \cdots, n - 1\}$, ${\alpha} \in \{0,1\}$, $x_j^1 = {x_j}$, and $x_j^0 = {{\overline {x}}_j}$.

Variable ${x_i}$ may have no or multiple variable mappings in the transformation detection procedure in this paper. If a mapping relation exists between two variables, then they must have the same $1^{st}$ signature value without consideration of order. A mapping exists between variables ${x_i}$ and ${x_j}$ in the following two cases, and the $1^{st}$ signature values of $x_i$ and $x_j$ are the same when they satisfy case (1) or (2).

(1) $(\left| {{f_{{x_i}}}} \right|,\left| {{f_{{{\overline {x}}_i} }}} \right|) = (\left| {{f_{{x_j}}}} \right|,\left| {{f_{{{\overline {x}}_j} }}} \right|)$

(2) $(\left| {{f_{{x_i}}}} \right|,\left| {{f_{{{\overline {x}}_i} }}} \right|) = (\left| {{f_{{{\overline {x}}_j} }}} \right|,\left| {{f_{{x_j}}}} \right|)$

When ${x_i}$ and ${x_j}$ satisfy only case (1), the two variables have a mapping with an identical phase, which is ${{\varphi _i}:x_i \to x_j }({\overline{x}}_i \to {\overline{x}}_j )$. We express this variable mapping as $i \to j-0$. When $x_i$ and $x_j$ satisfy only case (2), they have a mapping with an opposite phase, which is ${{\varphi _i}:x_i \to {{\overline {x}}_j} }({\overline{x}}_i \to x_j)$. We express this variable mapping as $i \to j-1$. If $x_i$ and $x_j$ satisfy both cases (1) and (2), then we must consider two variable mappings. In contrast, when $x_i$ and $x_j$ do not satisfy either case (1) or case (2), they have no variable mapping. In this paper, $i \to j - k:i,j \in \{0,1, \cdots, n - 1\},k \in \{0,1\}$ is used to express the variable mapping of variable ${x_i}$.

\emph{Example 3:} Consider the Boolean functions $f$ and $g$ from Example 2 with the same $1^{st}$ signature vector. Variable $x{}_1$ of $f$ and variable ${x_0}$ of $g$ satisfy case (2), with the variable mapping $1 \to 0-1$. Variables ${x_0}$ of $f$ and ${x_1}$ of $g$ satisfy both cases (1) and (2); therefore, two possible mappings exist: $0 \to 1-1$ and $0 \to 1-0$.

\emph{Definition 7 (Variable-Mapping Set):} Every variable has zero or more variable mappings. The variable-mapping set of ${x_i}$ is the collection that includes all possible variable mappings of ${x_i}$. The variable-mapping set of ${x_i}$ is denoted as ${\chi _i} = \{ {\varphi _i}:{x_i} \to x_j^{{\alpha}_j}|i,j \in \{0,1, \cdots, n - 1\},{\alpha}_j\in \{0,1\}\}$. In this paper, the variable-mapping set of ${x_i}$ is simplified to ${\chi _i} = \left\{ {i \to j - k|i,j \in \{0,1, \cdots, n - 1\},k \in \{0,1\}} \right\}$.

 The cardinality of the variable-mapping set with respect to $x_i$ is denoted as $\left| {{\chi _i}} \right|$. When $\left| {{\chi _i}} \right| = 1$, the variable-mapping set of $x_i$ is called a single-mapping set, and when $\left| {{\chi _i}} \right| > 1$, it is called a multiple-mapping set. In Example 3, the variable-mapping set of ${x_0}$ is a multiple-mapping set because ${\chi _0} = \left\{ {0 \to 1 - 0, 0 \to 1 - 1, 0 \to 2 - 0, 0 \to 2 - 1} \right\}$.

\emph{Definition 8 (Minimum Variable-Mapping Set):} An $n$-variable Boolean function has $n$ variable-mapping sets. The minimum variable-mapping set has the smallest cardinality.

 In Example 3, the cardinality of the variable-mapping sets of variables ${x_0}$  and ${x_2}$ in $f$ is 4, and the cardinality of the variable-mapping set of ${x_1}$ is 1. Therefore, the minimum variable-mapping set of function $f$ is ${\chi _1} = \left\{ {1 \to 0 - 1} \right\}$.

\emph{Definition 9 (Variable Symmetry):} If an $n$-variable Boolean function $f$  is invariant when variables $x{}_i$ and ${x_j}({{\overline {x}}_j}) $ are swapped, then variables $x{}_i$ and ${x_j}({{\overline {x}}_j}) $ are symmetric~\cite{26}.

Boolean functions can have skew and nonskew symmetries~\cite{26}. In this paper, we consider only nonskew symmetry, which reduces the complexity of Boolean matching because discovering this symmetry is inexpensive. Given a pair of variables $x{}_i$ and $x{}_j$, their cofactors can be used to check the symmetry. If ${f_{{x_i}{{\overline {x}}_j} }} \oplus {f_{{{\overline {x}}_i} {x_j}}} = 0$ or ${f_{{x_i}{x_j}}} \oplus {f_{{{\overline {x}}_i} {{\overline {x}}_j} }} = 0$, then variables $x{}_i$ and ${x_j}({{\overline {x}}_j})$ are symmetric.

We use this method in our algorithm to discover variable symmetry. When the algorithm detects symmetry, it checks only the variables that have the same $1^{st}$ signature value. In this paper, we classify variables into two types: symmetric and asymmetric. NP transformation does not change the symmetry property of a variable.

A symmetry class has more than one variable, and any two variables are symmetric. We use $C_i$ to denote a symmetry class and the first symmetry variable of $C_i$ is $x_i$. The number of variables in $C_i$ is denoted by its cardinality $\left| {{C_i}} \right|$.

\emph{Definition 10 (Symmetry Mapping):} Symmetry mapping is the correspondence relation between two symmetry classes of two Boolean functions. The symmetry mapping of symmetry class $C_i$ is denoted as ${\psi _i}:{C_i} \to {C_j}$, where $i,j \in \{0,1, \cdots, n - 2\}$. The symmetry mapping between $C_i$ and $C_j$ is abbreviated as ${i \to j}$.

Suppose that two NP-equivalent Boolean functions $f$ and $g$ and an NP transformation $T$ with a symmetry mapping $i1\to j1$ exist. If the symmetric variables of $C_{i1}$ of $f$ are $x_{i1}$ and $x_{i2}$ and the symmetric variables of $C_{j1}$ of $g$ are $x_{j1}$ and $x_{j2}$, then two variable mapping relations exist between $C_{i1}$ and $C_{j1}$: $\{i1-j1-k1,i2-j2-k2\}$ and $\{i1-j2-k3,i2-j1-k4\}$, where $k1,\cdots,k4 \in \{0,1\}$.

\emph{Definition 11 (Symmetry-Mapping Set):} Each symmetry class may have zero or more symmetry mappings. We define the symmetry-mapping set with respect to symmetry class $C_i$ as ${S_i} = \{ {\psi _i}:{C_i} \to {C_j}|i,j \in \{0,1, \cdots, n - 2\}\} $. In this paper, we simplify the symmetry-mapping set to ${S_i} = \{ i \to j|i,j \in \{0,1, \cdots n - 2\}\}$.

Similar to a variable-mapping set, a symmetry-mapping set whose cardinality is one, i.e., $\left| {{S_i}} \right| = 1$, is called a single symmetry-mapping set. When $\left| {{S_i}} \right| > 1$, $S_i$ is called a multiple symmetry-mapping set.

In our algorithm, we group all the variables. The basic principle of grouping is that variables having the same $1^{st}$ signature value belong to one group. According to the size of the $1^{st}$ signature value, we assign a serial number to each group.

\emph{Definition 12 (Structural Signature Vector):} An $n$-variable Boolean function $f$ has an SS vector ${V_f} = \{ {V_0},{V_1}, \cdots, {V_{n - 1}}\}$. ${V_i}$ is the structural signature value of ${x_i}$.

An SS vector is a new signature vector that adds structural information to the $1^{st}$ signature vector. The structural signature value of $x_i$ is ${V_i} = (\left| {{f_{{x_i}}}} \right|,\left| {{f_{{{\overline {x}}_i} }}} \right|,\left| {{C_i}} \right|,{C_i},{G_i})$. It includes a positive variable cofactor, a negative variable cofactor, the cardinality of the symmetry class to which $x_i$ belongs, the serial number of the first symmetric variable in its symmetry class, and a group serial number.

\emph{Example 4:} Consider the following two Boolean functions $f$ and $g$:

$f(X) = {{\overline {x}}_0} {{\overline {x}}_1} {x_3} + {{\overline {x}}_0} {x_1}{{\overline {x}}_3}  + {x_0}{{\overline {x}}_1} {{\overline {x}}_2} {x_3} + {x_0}{{\overline {x}}_1} {x_2}{{\overline {x}}_3}  + {x_0}{x_1}{x_3}$

$g(X) = {{\overline {x}}_0} {{\overline {x}}_1} {{\overline {x}}_3}  + {{\overline {x}}_0} {x_1}{{\overline {x}}_2} {x_3} + {{\overline {x}}_0} {x_1}{x_2} + {x_0}{{\overline {x}}_1} {x_3} + {x_0}{x_1}{{\overline {x}}_2} {{\overline {x}}_3} $
The SS vectors of $f$ and $g$ are as follows:

${V_f}$=\{(4, 4, 2, 0, 1),(4, 4, 2, 0, 1),(4, 4, -1, -1, 1),(5, 3, -1, -1, 0)\}

${V_g}$=\{(3, 5, -1, -1, 0),(4, 4, 2, 1, 1),(4, 4, -1, -1, 1),(4, 4, 2, 1, 1)\}

From the SS vectors of $f$ and $g$, we can obtain the following information:

(1) One symmetry class $C_0=\{x_0,x_1\}$ and two groups ${G_0} = \{ x_3\} $ and ${G_1} = \{ {x_0},{x_1},{x_2}\} $ exist in $f$.

(2) One symmetry class $C_1=\{x_1,x_3\}$ and two groups ${G_0} = \{ x_0\} $ and ${G_1} = \{ {x_1},{x_2},{x_3}\} $ exist in $g$.

(3) One single-mapping set ${\chi}_3=\{3\to0-1\}$, one multiple symmetry-mapping set $S_i=\{0\to1\}$ and one multiple-mapping set ${\chi}_2=\{2\to2-0,2\to2-1\}$ exist.

The symmetry-mapping set $S_0$ is a multiple symmetry-mapping set because the phases of variables in symmetry class $C_0$ of $f$ and $C_1$ of $g$ are not determined. In this case, we also consider the cases of identical and opposite phases.

\section{Matching Algorithm}
The core idea of the algorithm in this paper is to use an SS vector and Shannon decomposition to search for variable mappings and to form possible transformations. The algorithm is based on a fundamental strategy: recursive decomposition and searching. In this section, we describe the entire Boolean matching process.
\subsection{SS Vector Updating}

In the matching process, the phase of variable ${x_i}$ is obtained by comparing the relation between $\left| {{f_{{x_i}}}} \right|$ and $\left| {{f_{{{\overline {x}}_i} }}} \right|$. This method is similar to that in \cite{3}. If $\left| {{f_{{x_i}}}} \right| > \left| {{f_{{{\overline {x}}_i} }}} \right|$, then the phase of ${x_i}$ is positive. If $\left| {{f_{{x_i}}}} \right| < \left| {{f_{{{\overline {x}}_i} }}} \right|$, then the phase of ${x_i}$ is negative. If $\left| {{f_{{x_i}}}} \right| = \left| {{f_{{{\overline {x}}_i} }}} \right|$, then the phase of ${x_i}$ is undetermined. We use 0, 1, and -1 to denote a positive, negative and undetermined phase, respectively. In Example 4, the phase set of the variables of $f$ is Phase\_f=\{-1, -1, -1, 0\}. The phase set of the variables of $g$ is Phase\_g=\{1, -1, -1, -1\}.

Our algorithm will create a variable mapping between $x_i$ of Boolean function $f$ and $x_j$ of Boolean function $g$ if these two variables have the same SS value. However, if the phases of $x_i$ and $x_j$ are not determined, then we must consider two variable mappings $i\to j-0$ and $i\to j-1$. In Example 4, variable $x_2$ has two possible mappings $2\to2-0$ and $2\to2-1$ because its phase is not determined. To reduce the search space, we should determine the phases of variables as much as possible.

If the $1^{st}$ signature value of variable $x_2$ of Boolean functions $f$ and $g$ in Example 4 can be changed, then the phases of them can be determined; thus, the above problem is solved.

If the variable mapping set of $x_i$ of Boolean function $f$ is a multiple-mapping set, then multiple variables of $g$ with the same SS value as that of $x_i$ exist, and we must attempt multiple variable mappings for $x_i$. The existence of the multiple-mapping set is the cause of the big search space. Therefore, we should reduce the number of variable mappings in a multiple-mapping set.

The more variables that have the same SS value and undetermined phases, the larger the search space required to perform Boolean matching. Therefore, we utilize the identified variables to update the SS vector, which may change the SS vector and cause unidentified variables to have different SS values. Thus, updating the SS vector is an important step.

Consider two NP-equivalent Boolean functions $f$ and $g$ and the NP transformation $T$ with the variable mapping $x_i \to x_j$ between them. If we decompose $f$ using $x_i$ and decompose $g$ using $x_j$ via Shannon decomposition, then $f=x_if_{x_i}+\overline{x}_if_{\overline{x}_i}$ and $g=x_jg_{x_j}+\overline{x}_jg_{\overline{x}_j}$. Because the Boolean functions $f$ and $g$ are NP-equivalent under the NP transformation T, the Boolean functions $f_1=x_if_{x_i}$ and $g_1=x_jg_{x_j}$ must also be NP-equivalent, as well as $f_0=\overline{x}_if_{\overline{x}_i}$ and $g_0=\overline{x}_jg_{\overline{x}_j}$. Here, we refer to variables $x_i$ and $x_j$ as splitting variables. After two or more decompositions, more than one splitting variable exists. $cube\_f$ and $cube\_g$ are two cubes formed by splitting variables.

Since $f_1$ is NP-equivalent to $g_1$, we update the SS vector with them. The new SS vectors may be changed, and the phases of variables whose phases could not be determined previously may be determined now. Variables that had the same SS values previously may have different SS values after the update.

 A variable whose phase and variable mapping are determined is called an identified variable. Otherwise, it is an unidentified variable. In Example 4, the phase and variable mapping of variable $x_3$ of Boolean function $f$ are determined. Variable $x_3$ is an identified variable. The first variable mapping to be addressed is $3\to 0-1$. $f$ is decomposed using $x_3$, and $g$ is decomposed using ${\overline{x}}_0$, i.e., $cube\_f=x_3$ and $cube\_g={\overline{x}}_0$. The new SS vectors are computed by $x_3f_{x_3}$ and ${\overline{x}}_0g_{{\overline{x}}_0}$. Our algorithm uses $cube\_f$ and $cube\_g$ to update the two SS vectors. In the process of updating SS vectors, the $1^{st}$ signature value of already identified variables is set to (0,0). The new SS vectors of Example 4, updated using $x_3$ and ${\overline{x}}_0$, are as follows: ${V_f}$=\{(3, 2, 2, 0, 1),(2, 3, 2, 0, 1),(2, 3, -1, -1, 1),(0, 0, -1, -1, 0)\} and ${V_g}$=\{ (0, 0, -1, -1, 0),(3, 2, 2, 1, 1),(3, 2, -1, -1, 1),(2, 3, 2, 1, 1)\}. From these updated results, we know that the phases of all unidentified variables are determined and that the variable mapping set of $x_2$ of Boolean function $f$ has become a single-mapping set. Therefore, updating the SS vector is very useful for phase assignment and searching for variable mappings.

 Procedure 1 is carried out to update the SS vector of two Boolean functions $f$ and $g$.

\floatname{algorithm}{Procedure}
\renewcommand{\algorithmicrequire}{\textbf{Input:}}
\renewcommand{\algorithmicensure}{\textbf{Output:}}
    \begin{algorithm}
    \small
        \caption{SS Vector Updating}
        \begin{algorithmic}
            \Require  $f$, $g$, $cube\_f$, $cube\_g$,$V_f$,$V_g$
            \Ensure 0 or 1
             \Function {update}{$f, g, cube\_f, cube\_g$}
                \State Compute\_vector($f, cube\_f$)
                \State Compute\_vector($g, cube\_g$)
                \State Group($V_f$)
                \State Group($V_g$)
                \State Update the phase of variables of $f$
                \State Update the phase of variables of $g$
                \State Return vector\_same($V_f, V_g$)
            \EndFunction
        \end{algorithmic}
    \end{algorithm}

\subsection{Searching Variable Mappings}

To perform NPN matching of two $n$-variable Boolean functions, each transformation has $n$ variable mappings. Searching for variable mappings is critical. In Section \uppercase\expandafter{\romannumeral3}, variables are classified as either symmetric or asymmetric. The variable mapping set of an asymmetric variable is classified as either a single-mapping set or a multiple-mapping set. The variable mapping set of a symmetric variable is classified as either a single symmetry-mapping set or a multiple symmetry-mapping set.

 The elementary principle of a variable mapping existing between two variables is that they have the same SS values. In the comparison of two SS values, two variables have a mapping if they have the same $1^{st}$ signature value, number of variables in the symmetry class and group number. A variable mapping is created among asymmetric variables or among symmetric variables.

 Our algorithm creates a variable mapping $i-j-k$ directly if the variable mapping set of $x_i$ of $f$ is a single-mapping set and only the variable $x_j$ of $g({\overline{g}})$ has the same SS value as that of $x_i$. If the variable mapping set of variable $x_i$ of $f$ is a multiple-mapping set ${\chi}_i=\{i\to j1-k1,\cdots,i\to jm-km\}$, where $2 \ge m\le 2n$ and $k1,\cdots,km \in \{0,1\}$, our algorithm first creates the variable mapping from $x_i$ of $f$ to $x_{j1}$ of $g({\overline{g}})$. If the transformations $T$ generated by the first variable mapping of ${\chi}_i$ do not satisfy $f(TX)=g(X)({\overline{g(X)}})$, our algorithm will return and fetch the next variable mapping from ${\chi}_i$. The searching operation is terminated when a transformation $T$ that satisfies $f(TX)=g(X)({\overline{g(X)}})$ is generated by ${\chi}_i$ or when none of the transformations $T$ generated by ${\chi}_i$ satisfy $f(TX)=g(X)({\overline{g(X)}})$.

 When the variable $x_i$ of $f$ is a symmetric variable and its variable mapping set is a single symmetry-mapping set $S_i=\{i\to j\}$ and $C_i=\{x_{i},x_{i1},\cdots,x_{im}\}$,$C_j=\{x_{j},x_{j1},\cdots,x_{jm}\}$, where $1 \ge m \le n-1$, each variable of $C_{i}$ can be mapped to any variable of $C_{j}$. Thus, there are $m!$ possible variable mapping relations between $C_{i}$ and $C_{j}$. According to the invariant property of swapping symmetric variables, these $m!$ variable mapping relations are equivalent. Therefore, we do not address every variable mapping relation between $C_i$ and $C_j$. Consequently, our algorithm establishes only one variable mapping relation, which has $m$ variable mappings $\{i\to j-k,i1\to j1-k1,\cdots,im\to jm-km\}$, where $1 \ge m \le n-1$ and $k,k1,\cdots\,km \in \{0,1\}$. If the variable mapping set of symmetric variable $x_i$ is a multiple symmetry-mapping set $S_i=\{i\to j1,\cdots,i\to jm\}$, where $2 \ge m \le \left\lfloor {n/2} \right\rfloor$, then our algorithm first addresses the variable mapping relation generated by $i\to{j1}$. If the transformations $T$ that are generated by $i\to j1$ do not satisfy $f(TX)=g(X)({\overline{g(X)}})$, then our algorithm will return and fetch the next symmetry mapping. The searching operation is terminated when a transformation $T$ that satisfies $f(TX)=g(X)({\overline{g(X)}})$ is generated by $S_i$ or when none of the transformations $T$ generated by $S_i$ satisfy $f(TX)=g(X)({\overline{g(X)}})$.

Based on the above discussion, we test all possible variable mappings between Boolean functions $f$ and $g$. Therefore, our algorithm must find a transformation $T$ that satisfies $f(TX)=g(X)({\overline{g(X)}})$ if Boolean function $f$ is NPN-equivalent to $g$. If Boolean function $f$ is not NPN-equivalent to $g$, then either a transformation $T$ may not be generated or none of the generated transformations $T$ satisfy $f(TX)=g(X)({\overline{g(X)}})$.

In Example 4, after the first SS vector update, one single-mapping set ${\chi}_2=\{2\to 2-1\}$ and one single symmetry-mapping set $S_0=\{0\to 1\}$ exist. Our algorithm creates variable mappings $0\to 1-0,1\to 3-0$ and $2\to 2-1$.

The use of symmetry reduces the search space considered during matching. The group mark also reduces the search space to some extent. If two Boolean functions $f$ and $g$ are NP equivalent and the variable $x_i$ of $f$ is permuted to $x_j({\overline{x}}_j)$ of $g$, the change of the $1^{st}$ signature value of the variable $x_i$ and that of $x_j$ must be synchronous.

In the matching process for two Boolean functions $f$ and $g$, two cases exist that will produce disadvantages if there is no group mark in the SS vector. These two cases are called the $1^{st}$ signature value collisions.

(1) Some variables have the same $1^{st}$ signature values in the previous recursion but have different $1^{st}$ signature values in the subsequent recursion.

(2) Some variables have different $1^{st}$ signature values in the previous recursion but have the same $1^{st}$ signature values in the subsequent recursion.

In case 1, if our algorithm creates variable mappings between these variables in the previous recursion, then the transformations generated by these variable mappings must be incorrect. If we create variable mappings in the subsequent recursion in case 2, then the transformations generated by these variable mappings must be incorrect.

To reduce the disadvantages produced by case 1 and case 2, our algorithm first addresses all single-mapping sets and single symmetry-mapping sets in each recursion. If single-mapping sets or single symmetry-mapping sets exist, then our algorithm creates all variable mappings generated by them and enters the next recursion. Our algorithm selects the minimum variable-mapping set if no single-mapping set or single symmetry-mapping set exists. Each recursion updates the two SS vectors, $cube\_f$ and $cube\_g$. Updating the SS vector may lead to a change in the SS vectors and put these unidentified variables into different groups.

\emph{Example 5:} Assume that we are given two 5-variable Boolean functions:
$f(X)={\overline{x}}_0{\overline{x}}_1x_2x_4+x_1x_2{\overline{x}}_4+x_0{\overline{x}}_1x_3+x_0x_1{\overline{x}}_3x_4+x_0{\overline{x}}_1x_2{\overline{x}}_4
+x_0{\overline{x}}_1{\overline{x}}_2{\overline{x}}_3x_4+{\overline{x}}_0x_1{\overline{x}}_2x_3x_4  $

$g(X)={\overline{x}}_0x_1x_4+{\overline{x}}_1x_2{\overline{x}}_3{\overline{x}}_4+{\overline{x}}_0{\overline{x}}_1x_2x_3+x_0x_1x_2x_3+x_0{\overline{x}}_1x_2{\overline{x}}_3
+{\overline{x}}_0{\overline{x}}_1{\overline{x}}_2{\overline{x}}_3x_4+{\overline{x}}_0{\overline{x}}_2x_3{\overline{x}}_4+{\overline{x}}_0x_1x_2{\overline{x}}_3
+x_0{\overline{x}}_1{\overline{x}}_2x_3x_4  $

(1) In the first recursive call, $cube\_f=bddtrue$ and $cube\_g=bddtrue$, and the SS vectors of the two Boolean functions are as follows:

$V_f$=\{(11, 5, -1, -1, 0),(8, 8, -1, -1, 3),(10, 6, -1, -1, 1),(9, 7, -1, -1, 2),(9, 7, -1, -1, 2)\}

$V_g$=\{(5, 11, -1, -1, 0),(8, 8, -1, -1, 3),(10, 6, -1, -1, 1),(9, 7, -1, -1, 2),(9, 7, -1, -1, 2)\}

The two SS vectors are the same, and we obtain the variable phases of $f$ and $g$, which are Phase\_f=\{0, -1, 0, 0, 0\} and Phase\_g=\{1, -1, 0, 0, 0\}. At this point, the phases of variable $x_1$ of $f$ and $g$ are undetermined. The $1^{st}$ signature value of variables $x_1$ of $f$ is different with that of $x_3$ and $x_4$ of $g$.

 There are two single-mapping sets: ${\chi _0} = \{ 0\to 0-1\}$ and ${\chi _2} = \{ 2\to 2-0\}$. Our algorithm creates two variable mappings: $0\to 0-1$ and $2\to 2-0$. Then, $cube\_f$ is updated to $x_0$, and $cube\_g$ is updated to ${\overline{x}}_0$. Our algorithm enters the next recursion and updates the SS vectors.

(2) In the second recursive step, our algorithm utilizes ${x_0}$ and ${{\overline {x}}_0}$ to update the SS vectors. The updating results are as follows:

$V_f$=\{(0, 0, -1, -1, 0),(5, 6, -1, -1, 3),(0, 0, -1, -1, 1),(6, 5, -1, -1, 2),(6, 5, -1, -1, 2)\}

$V_g$=\{(0, 0, -1, -1, 0),(6, 5, -1, -1, 3),(0, 0, -1, -1, 1),(6, 5, -1, -1, 2),(6, 5, -1, -1, 2)\}

$Phase\_f=\{0,1,0,0,0,\}$, $Phase\_g=\{1,0,0,0,0\}$

From the updated SS vectors, we can obtain the following information:

1) The phases of all variables are determined.

2) The changed SS vectors are conducive to variable identification. We can identify variable ${x_1}$ of $f$ and $g$. A single-mapping set ${\chi _1}=\{1 \to 1-1\}$ exists. Our algorithm updates $cube\_f={x_0}{x_2}$ and $cube\_g={{\overline{x}}_0}{x_2}$.

In the first computation of SS vectors, the variable $x_1$ of $f$ has a different $1^{st}$ signature value compared to that of variables $x_3$ and $x_4$ of $g$. However, their $1^{st}$ signature values become same during the second recursion. If no group mark exists, the variable mapping set of $x_1$ of $f$ is ${\chi}_1=\{1\to 1-1, 1\to 3-1, 1\to 4-1\}$. The variable mappings $1\to 3-1$ and $1\to 4-1$ are incorrect mappings. The use of a group mark reduces the probability of producing incorrect variable mappings. Therefore, it also reduces the search space.

\subsection{Transformation Detection}

The main goal of our algorithm is to create all possible NP transformations $T$ that map $f$ to $g(\overline g)$ and to verify whether a $T$ exists that satisfies $f(TX)=g(X)({\overline{g(X)}})$. An $n$-variable Boolean function has $n!{2^n}$ NP transformations, which results in a very high complexity. However, in reality, the number of NP transformations that can transform $f$ into $g(\overline g)$ is small. Therefore, it is very important to find those incorrect transformations as soon as possible.

Two strategies for finding the incorrect transformations are:

(1) Compare two updated SS vectors

If the Boolean functions $f$ and $g({\overline{g}})$ are NPN-equivalent and a variable mapping $i-j-k$ exists, then the corresponding two Boolean functions (after decomposition) must have the same SS vectors. In contrast, the variable mapping $i-j-k$ must be incorrect if the updated SS vectors are not the same.

(2) Phase collision check

Phase collision is a change in the phase relation between two variables. When a multiple-mapping set or a multiple symmetry-mapping set is found, multiple variable mappings or symmetry mappings can be selected. Our algorithm selects the variable mapping or symmetry mapping in sequence. Incorrect variable mappings or symmetry mappings may lead to phase collision.

 Suppose that the minimum variable-mapping set is a multiple-mapping set ${\chi _i} = \left\{ {i \to {j_1} - {k_1}, \cdots, i \to {j_m} - {k_m}} \right\}$. Our algorithm first selects $i \to {j_1} - {k_1}$. Assuming that variables ${x_h}$ and ${x_l}$ have the same phase, our algorithm updates $cube\_f$ and $cube\_g$ and then calls a recursion.

In subsequent recursions, our algorithm updates the SS vectors and searches for variable mappings. If we find a variable mapping $h \to l - 1$, then ${x_h}$ and ${x_l}$ have opposite phases. A phase collision appears, as these two variables have the same phases as in the previous recursion. Phase collision is caused by the incorrect variable mapping $i \to {j_1} - {k_1}$.

Procedure 2 realizes the function of transformation detection. In the process of transformation detection, Procedure 2 creates a tree to build and store all possible transformations. Each unabridged branch is a transformation $T$. This tree has $n$ layers, and each layer has one or more variable mapping nodes. Procedure 2 utilizes DFS to detect a transformation. As long as a transformation $T$ is generated, verify() is called to determine whether $f(TX) = g(X)$ or $f(TX) = \overline{g(X)}$. Procedure 2 is terminated until either a transformation $T$ that can satisfy this condition is found or until no transformations can satisfy this condition.

\floatname{procedure}{procedure}
\renewcommand{\algorithmicrequire}{\textbf{Input:}}
\renewcommand{\algorithmicensure}{\textbf{Output:}}
    \begin{algorithm}
    \small
        \caption{Transformation Detection}
        \begin{algorithmic} 
            \Require $f, g, cube\_f, cube\_g, map\_list,V_f,V_g$
            \Ensure 0 or 1
            \Function {detect}{$f, g, cube\_f, cube\_g, map\_list,V_f,V_g$}
                 \If {$D_1$}
                    \State Create a transformation $T$
                    \State Return VERIFY($f,g,map\_list$)
                 \ElsIf {UPDATE($f, g, cube\_f, cube\_g$)=0}
                     \State Return 0
                 \Else
                    \ForAll {$x_i \in f(x)$}
                     \State Compute ${\chi}_i/S_i$
                        \If {$| {{\chi _i}}| = 1$ or $| {{S_i}} | = 1  $}
                           \State Create Map\_node for $x_i$ or $C_i$
                            \If {$D_2$}
                               \State Return 0
                            \EndIf
                            \State Add Map\_node to Map\_list
                         \Else
                            \State Get  $Min(\left| {{\chi _i}} \right|)$
                         \EndIf
                     \EndFor
                     \If {$D_3$}
                         \State Update $cube\_f$ and $cube\_g$
                         \State Return DETECT($f, g, cube\_f, cube\_g, map\_list$)
                     \Else
                           \State Create Map\_node from ${\chi}_{min}$ or $S_{min}$
                            \ForAll {$ Node \in Map\_node$}
                                \If {$D_2$}
                                    \State Continue
                                \Else
                                    \State Add node to Map\_list
                                    \State Update $cube\_f$ and $cube\_g$
                                    \State Return DETECT($f, g, cube\_f, cube\_g, map\_list$
                                \EndIf
                           \EndFor
                           \State Return 0
                      \EndIf
                  \EndIf
            \EndFunction
        \end{algorithmic}
    \end{algorithm}

In Procedure 2, Map\_node represents one or more variable mappings. Map\_list is the transformation tree. Conditions $D_1$, $D_2$ and $D_3$ are as follows:

$D_1$ occurs when a branch of the transformation tree exists whose layers reach $n$.

$D_2$ occurs when the current variable mapping has a phase collision.

$D_3$ occurs when the minimum variable-mapping set is a single-mapping set or a single symmetry-mapping set.

During each recursion, Procedure 2 identifies one or more variables and adds one or more variable mapping nodes to the transformation tree. When more than one single-mapping set or single symmetry-mapping set exists, the depth of the transformation tree is increased more than once. In each recursion, Procedure 2 updates $cube\_f$ and $cube\_g$.

When a layer of the transformation tree has more than one node, incorrect variable mappings may exist. These incorrect mappings generate incorrect branches, which increase the search pace and slow down matching. Therefore, we must find incorrect variable mappings and prune these branches. Procedure 2 prunes a branch when the two updated SS vectors are not equal or when a new variable mapping introduces a phase collision.

\emph{Example 6:} For the Boolean functions in Example 5, after the first decomposition and updating of SS vectors, three variables are already identified. Variable $x_4$ of $f$ and that of $g$ have the same phase. Now, $cube\_f={x_0}{x_2}$ and $cube\_g={{\overline{x}}_0}{x_2}$.

(1) Because condition $D_1$ is not satisfied, the SS vectors are updated. The results are as follows:

$V_f$=\{(0, 0, -1, -1, 0),(0, 0, -1, -1, 3),(0, 0, -1, -1, 1),(3, 3, -1, -1, 2),(3, 3, -1, -1, 2)\}

$V_g$=\{(0, 0, -1, -1, 0),(0, 0, -1, -1, 3),(0, 0, -1, -1, 1),(3, 3, -1, -1, 2),(3, 3, -1, -1, 2)\}

The two SS vectors above have two minimum variable-mapping sets: ${\chi _3}=\{3\to 3-0,3\to 4-0\}$ and ${\chi _4}=\{4 \to 3-0,4 \to 4-0\}$. Procedure 2 selects the first minimum variable set $\chi _3$ to address. The transformation tree is extended by one layer, with two variable nodes $3\to3-0$ and $3\to4-0$. This minimum mapping set has two variable mappings. Procedure 2 selects $3\to3-0$ and then updates $cube\_f$ and $cube\_g$ with $cube\_f={x_0}{x_2}{{\overline{x}}_1}$ and $cube\_g={{\overline{x}}_0}{x_2}{x_1}$.

 (2) Procedure 1 updates the SS vectors. The results are as follows:

$V_f$=\{(0, 0, -1, -1, 0),(0, 0, -1, -1, 3),(0, 0, -1, -1, 1),(0, 0, -1, -1, 2),(1, 2, -1, -1, 2)\}

$V_g$=\{(0, 0, -1, -1, 0),(0, 0, -1, -1, 3),(0, 0, -1, -1, 1),(0, 0, -1, -1, 2),(2, 1, -1, -1, 2)\}

From the above SS vector results, Procedure 2 finds a single-mapping set ${\chi}_4=\{4\to 4-1\}$. The phase relations of variable $x_4$ of $f$ and that of $g$ are opposite. However, we know that they had the same phase previously; therefore, we find a phase collision. The reason this collision occurs is the selection of the variable mapping $3\to3-0$. Therefore, this branch is pruned. Procedure 2 returns $3\to4-0$ and continues to search the rest of the variable mappings.

From Example 6, we see that our algorithm prunes the incorrect branch in time. If we did not check for a phase collision, the wrong variable mapping might have yielded one or more incorrect transformations. Therefore, phase collision check reduces the search space and accelerates the algorithm to some extent.

\subsection{Boolean Matching}

Given two Boolean functions $f$ and $g$, we should first verify the phases of these two Boolean functions. In our algorithm, the phase of $f$ is positive, and the phase of $g$ is determined by the relation between the zeroth-order signature values. If $|f| = |g| \land |f| \ne |\overline{g}|$, then the phase of $g$ is positive. The phase of $g$ is negative when $|f| \ne |g| \land |f| = |\overline{g}|$. We may test both positive and negative cases when $|f| = |g| \land |f| = |\overline{g}|$.

Multiple transformations exist between two NPN-equivalent Boolean functions. We need to find only one transformation that can transform $f$ into $g(\overline{g})$ to prove that $f$ is equivalent to $g(\overline{g})$. We explained the process of transformation detection in part C. The two Boolean functions are equivalent when calling Matching() returns 1; otherwise, they are non-equivalent. Together, Procedures 1, 2 and 3 form the NPN Boolean matching algorithm of this paper.

\floatname{procedure}{procedure}
\renewcommand{\algorithmicrequire}{\textbf{Input:}}
\renewcommand{\algorithmicensure}{\textbf{Output:}}
    \begin{algorithm}
    \small
        \caption{Boolean Matching}
        \begin{algorithmic}
            \Require $f, g$
            \Ensure 0 or 1
            \Function {Matching}{$f, g$}
            \State $cube\_f=bddtrue$, $cube\_g=bddtrue$, $map\_list=NULL,V_f=NULL,V_g=NULL$
            \If {$|f|=|g|$}
              \If {$|f|\ne|g|$}
                  \If {DETECT($f, g, cube\_f, cube\_g, map\_list$)=1}
                      \State Return 1
                  \Else                                                 
                       \State Return 0
                  \EndIf
              \Else
                  \If {DETECT($f, g, cube\_f, cube\_g, map\_list$)=1}
                      \State Return 1
                  \Else                                                 
                      \State g=!g
                      \If {DETECT($f, g, cube\_f, cube\_g, map\_list$)=1}
                          \State Return 1
                      \Else                                                 
                         \State Return 0
                      \EndIf
                  \EndIf
              \EndIf
           \ElsIf {$|f|=|\overline{g}|$}
              \State g=!g
              \If {DETECT($f, g, cube\_f, cube\_g, map\_list$)=1}
                 \State Return 1
              \Else                                                 
                 \State Return 0
              \EndIf
           \Else
              \State Return 0
           \EndIf
            \EndFunction
        \end{algorithmic}
    \end{algorithm}

\emph{Example 7:} Consider the two 7-variable Boolean functions $f(X)$ and $g(X)$:

$f(X) = {\overline{x}}_0x_2x_5{\overline{x}}_6+{\overline{x}}_0x_1x_2{\overline{x}}_3x_6+x_1{\overline{x}}_2{\overline{x}}_3x_6+x_0x_4{\overline{x}}_5+x_0x_2x_5{\overline{x}}_6+x_0x_1x_2{\overline{x}}_3{\overline{x}}_4x_6+x_0x_1x_2{\overline{x}}_3x_4x_5x_6  $

$g(X)={\overline{x}}_0{\overline{x}}_1{\overline{x}}_2{\overline{x}}_5+{\overline{x}}_0{\overline{x}}_1{\overline{x}}_5x_6+{\overline{x}}_0{\overline{x}}_1{\overline{x}}_2x_5{\overline{x}}_6+
{\overline{x}}_0x_1x_3x_4+{\overline{x}}_0x_1{\overline{x}}_2x_5{\overline{x}}_6+{\overline{x}}_0x_1{\overline{x}}_2{\overline{x}}_5{\overline{x}}_6+
x_0x_1x_3x_4+x_0{\overline{x}}_1{\overline{x}}_5x_6   $

(1) The phases of $f$ and $g$ are positive. The initial value of both $cube\_f$ and $cube\_g$ is $bddtrue$. Condition $D_1$ is false. Procedure 1 computes the SS vectors of $f$ and $g$. The results are as follows:

${V_f}$=\{(30, 16, 2, 0, 1),(30, 16, 2, 1, 1),(31, 15, -1, -1, 0),(16, 30, 2, 1, 1),(30, 16, 2, 0, 1),(24, 22, -1, -1, 2),(22, 24, -1, -1, 2)\}

${V_g}$=\{(16, 30, 2, 0, 1),(22, 24, -1, -1, 2),(16, 30, 2, 0, 1),(30, 16, 2, 3, 1),(30, 16, 2, 3, 1),(15, 31, -1, -1, 0),(24, 22, -1, -1, 2)\}

The two SS vectors are the same. Procedure 2 searches a single-mapping set ${\chi _2} = \{ {2 \to 5 - 1}\}$ and adds variable mapping $2\to5-1$ to the transformation tree. $cube\_f$ and $cube\_g$ are updated with $cube\_f = {x_2}$ and $cube\_g = {{\overline {x}}_5}$. Then, the next recursive call occurs.

(2) Condition $D_1$ is false. Procedure 1 updates the SS vectors. The results are as follows:

 ${V_f}$=\{(19, 12, 2, 0, 1),(19, 12, 2, 1, 1),(0, 0, -1, -1, 0),(12, 19, 2, 1, 1),(19, 12, 2, 0, 1),(20, 11, -1, -1, 2),(11, 20, -1, -1, 2)\}

 ${V_g}$=\{(12, 19, 2, 0, 1),(11, 20, -1, -1, 2),(12, 19, 2, 0, 1),(19, 12, 2, 3, 1),(19, 12, 2, 3, 1),(0, 0, -1, -1, 0),(20, 11, -1, -1, 2)\}

These two SS vectors are the same. Procedure 2 searches through 4 variable mapping sets: $S_0=\{0\to0,0\to3\}$, $S_1=\{1\to0,1\to3\}$, ${\chi}_5=\{5\to 1-1,5\to 6-0\}$, and ${\chi}_6=\{6\to 1-0,6\to 6-1\}$. The cardinalities of these four variable mapping sets are all 2. Procedure 2 selects the first minimal variable-mapping set to address. $S_0$ is a multiple symmetry-mapping set. Two branches are generated by the symmetry mappings $0 \to 0$ and $0\to3$. Procedure 2 selects symmetry mapping $0\to0$ and generates two variable mappings $0 \to 0 - 1$ and $4 \to 2 - 1$. Then, it adds two layers to the transformation tree, each of which has one variable mapping node. If the transformations created by symmetry mapping $0\to0$ are verified to be false, Procedure 2 returns and selects $0\to3$. Then, Procedure 2 updates $cube\_f$ and $cube\_g$ with $x_2x_0$ and ${\overline{x}}_5{\overline{x}}_0$ and continues with the next recursive call.

(3) Condition $D_1$ is false. Procedure 1 updates the SS vectors, with the results being the following:

${V_f}$=\{(0, 0, 2, 0, 1),(11, 8, 2, 1, 1),(0, 0, -1, -1, 0),(8, 11, 2, 1, 1),(0, 0, 2, 0, 1),(10, 9, -1, -1, 3),(7, 12, -1, -1, 2)\}

${V_g}$=\{(0, 0, 2, 0, 1),(7, 12, -1, -1, 2),(0, 0, 2, 0, 1),(11, 8, 2, 3, 1),(11, 8, 2, 3, 1),(0, 0, -1, -1, 0),(10, 9, -1, -1, 3)\}

From the above updated SS vectors, we know that these two SS vectors are the same and that one single symmetry-mapping set ($S_1=\{1\to 3\}$) and two single-mapping sets (${\chi}_5=\{5\to 6-0\}$ and ${\chi}_6=\{6\to 1-0\}$) exist. Procedure 2 adds variable mappings $1\to 3-0$, $3\to 4-1$, $5\to 6-0$ and $6\to 1-0$ to the transformation tree. $Cube\_f$ is updated to $x_2x_0x_4$, and $Cube\_g$ is updated to ${\overline{x}}_5{\overline{x}}_0{\overline{x}}_2$. Procedure 2 then enters the next recursion.

(4) Condition $D_1$ is true. Procedure 2 creates a transformation $T = \{ 2 \to 5 - 1,0 \to 0 - 1,4 \to 2 - 1,1 \to 3 - 0,3 \to 4 - 1,5 \to 6 - 0,6 \to 1 - 0\} $.

During the process of transformation detection, each unabridged branch of the transformation tree is a detected NP transformation. Considering the Boolean matching of Example 7, $7!{2^7}$ possible transformations can be carried out using the exhaustive method. In contrast, using our proposed algorithm, only 2 possible transformations exist. All possible transformations are shown in Fig. 1.

\begin{figure}[!h]
\centering
\includegraphics[width=0.4\textwidth,height=5cm]{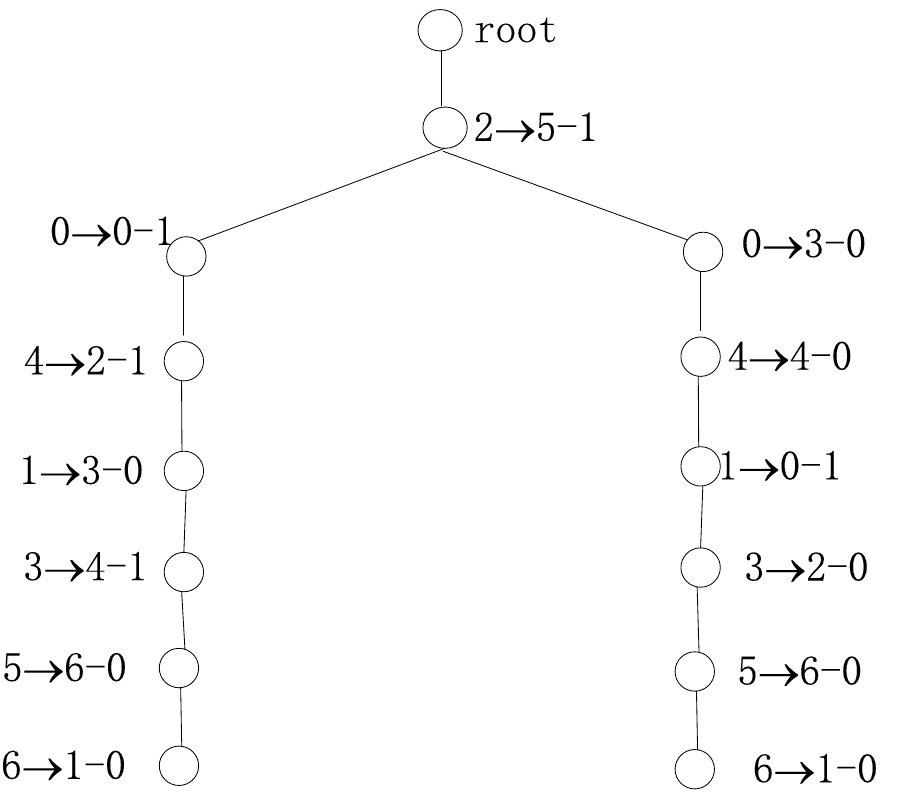}
\caption{Transformation Search Tree for Example 7}
\label{fig.1}
\end{figure}

Because Procedure 2 utilizes the DFS method, it may not be able to verify all the transformations in the transformation tree. In the execution of Example 7, our algorithm returns 1 when Procedure 2 detects the first transformation. Fig. 2 shows the actual transformation search tree of Example 7.

\begin{figure}[!h]
\centering
\includegraphics[width=0.3\textwidth,height=4cm]{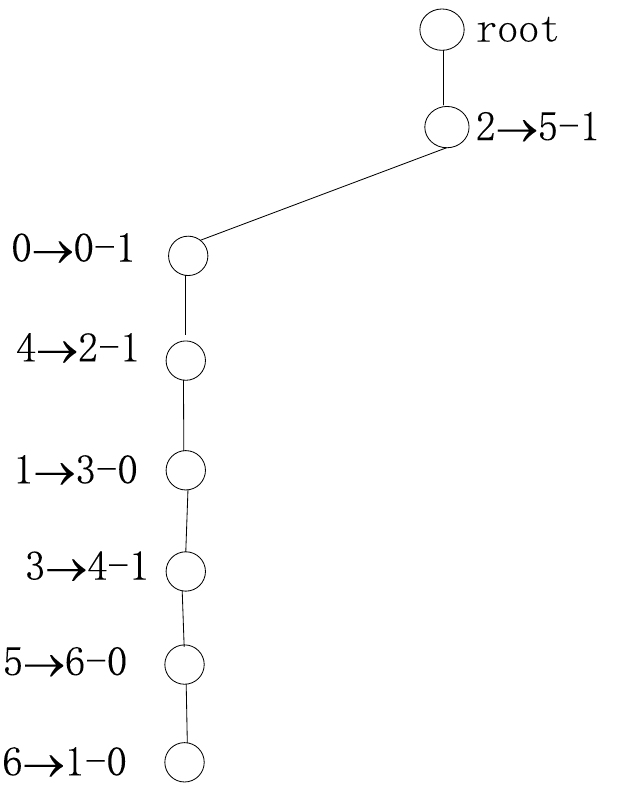}
\caption{The Actual Transformation Search Tree for Example 7}
\label{fig.2}
\end{figure}

From Example 7, use of the SS vector and Shannon expansion greatly reduces the search space. Use of the DFS method further improves the matching speed.

\section{Experimental Results}

To demonstrate the effectiveness of our proposed algorithm, we conducted an experiment using large circuit sets. We executed our algorithm using input sets containing NPN-equivalent and
 NPN non-equivalent circuits. We tested the algorithm using both MCNC benchmark circuits and randomly generated circuits.  All combinational circuits of the MCNC benchmark were tested. We use two different ways to generate random circuits. In the first way, the number of minterms of Boolean functions of the circuits are random. In the second way, the number of minterms of the Boolean functions of the circuits are $2^{n-1}$. For the non-equivalence testing experiment, the two Boolean functions $f$ and $g$ had the same or complementary zeroth-order signature values. We tested Boolean functions with 7-22 input variables.

In this paper, we compare the results of our algorithm with those of \cite{3}. The authors of \cite{3} proposed a canonical-based Boolean matching algorithm. Because \cite{3} did not report the runtimes for Boolean matching, we re-implemented the algorithm in \cite{3} and tested it on the Boolean function set generated from the MCNC benchmark to ensure that the results for computing canonical forms were consistent with those of \cite{3} within a comparable hardware environment.

Then, we compared these two algorithms within a new hardware environment with a 3.3 GHz CPU and 4 GB of RAM. The runtime we report is the CPU time used. The runtimes are reported in seconds and include Boolean functions with up to 22 inputs. We report the minimum, maximum and average runtimes of the algorithms proposed in this paper and the algorithm from \cite{3}. In the following tables, the first column lists the number of input variables (\#I), followed by three columns that show the minimum (\#MIN), maximum (\#MAX) and average  (\#AVG) runtimes of our algorithm. The next three columns are the corresponding runtimes of the algorithm from \cite{3}. The last columns of Tables \uppercase\expandafter{\romannumeral1}, \uppercase\expandafter{\romannumeral2} and \uppercase\expandafter{\romannumeral3} list the average BDD size (\#AVG nodes).

Table \uppercase\expandafter{\romannumeral1} shows the results for the equivalent MCNC benchmark circuits. Fig. 3 shows the average runtime of our algorithm compared to the average runtime of \cite{3} for the equivalent MCNC benchmark circuits.
\begin{table}[htp]
\label{Tab1}
\caption{Boolean matching runtimes on equivalent MCNC benchmark circuits}
\centering

\begin{tabular}{|p{0.2cm}<{\centering}|p{0.75cm}<{\centering}| p{0.8cm}<{\centering}|p{0.8cm}<{\centering}|p{0.8cm}<{\centering}| p{0.8cm}<{\centering}|p{0.8cm}<{\centering}|p{0.7cm}<{\centering}|}
\hline
\#I &\#MIN & \#MAX &\#AVG  &\#MIN of Ref\cite{3} & \#MAX of Ref\cite{3} & \#AVG of Ref\cite{3} & \#AVG nodes\\
\hline
7&0.00002 	&0.00035 	&0.00011 	&0.00004 	&0.05302 	&0.00121 &19108 \\
\hline
8&0.00013 	&0.00059 	&0.00030 	&0.00008 	&0.01164 	&0.00122 &30616 \\
\hline
9 &0.00005 	&0.00111 	&0.00044 	&0.00011 	&0.00318 	&0.00186 &41391\\
\hline
10 &0.00011 	&0.00188 	&0.00063 	&0.00020 	&0.00245 	&0.00193 &170847 \\
\hline
11&0.00009 	&0.00217 	&0.00079	&0.00027 	&0.00505 	&0.00243 &124842\\
\hline
12& 0.00018 	&0.00625 	&0.00090 	&0.00025 	&0.01480 	&0.00255 &282490 \\
\hline
13 & 0.00089 	&0.02500 	&0.00140 	&0.00126 	&0.19924 	&0.00535 &554213  \\
 \hline
14 & 0.00073 	&0.01809 	&0.00360 	&0.00112 	&0.03291 	&0.01245 &540672 \\
\hline
15 & 0.00039 	&0.05908 	&0.00478	&0.00102 	&0.14415 	&0.04077  &574421 \\
\hline
16& 0.00064 	&0.31375	&0.01232 	&0.00074 	&0.06281 	&0.04849  &604159 \\
\hline
17 & 0.00102 	&0.65791	&0.11360 	&0.00067 	&1.19085 	&0.31644  &545496\\
\hline
18 & 0.00192 	&1.63740 	&0.23490 	&0.00132 	&4.10735 	&0.64273  &560014\\
\hline
19 & 0.00391 	&1.94918	&0.88743	&0.18556 	&3.83515 	&1.46287  &508160\\
\hline
20 & 0.00904 	&4.75490 	&1.28150 	&0.18556 	&10.44000 	&2.13027  &562698\\
\hline
21 & 0.16699 	&5.74320 	&3.73066 	&0.29532 	&29.25150 	&10.21224 &523303   \\
\hline
22& 1.29078 	&16.44130 	&6.24368 	&1.24569 	&30.36820 	&11.27597 &621788 \\
\hline
\end{tabular}
\end{table}

\begin{figure}[!h]
\centering
\includegraphics[width=0.4\textwidth,height=7cm]{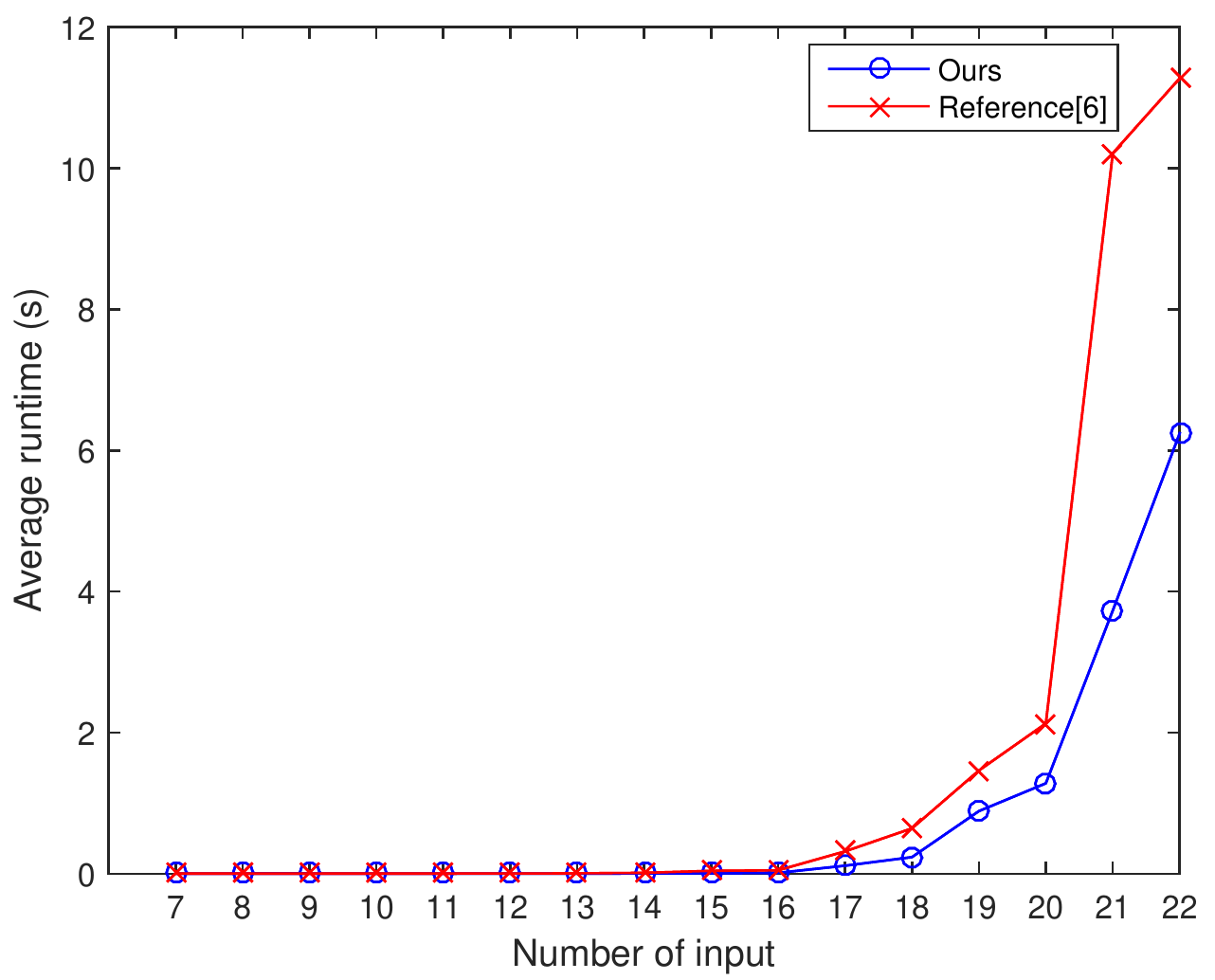}
\caption{Comparison of average runtimes on equivalent MCNC benchmark circuits}
\label{fig.3}
\end{figure}

From the results of Table \uppercase\expandafter{\romannumeral1}, we can see that the matching speed our algorithm is faster than the algorithm of \cite{3}. The average runtime of our algorithm is 3.8 times faster than competitors when testing the equivalent MCNC benchmark circuits.

Table \uppercase\expandafter{\romannumeral2} shows the results from testing the first type of equivalent random circuits, and Fig. 4 shows the average runtime of our algorithm compared with that of \cite{3} for the first type of equivalent random circuits.

\begin{table}[htp]
\label{Tab2}
\caption{Boolean matching runtimes on the first type of equivalent random circuits}
\centering

\begin{tabular}{|p{0.2cm}<{\centering}|p{0.75cm}<{\centering}| p{0.8cm}<{\centering}|p{0.8cm}<{\centering}|p{0.8cm}<{\centering}| p{0.8cm}<{\centering}|p{0.8cm}<{\centering}|p{0.7cm}<{\centering}|}
\hline
 \#I &\#MIN & \#MAX &\#AVG  &\#MIN of Ref\cite{3} & \#MAX of Ref\cite{3} & \#AVG of Ref\cite{3} & \#AVG nodes\\
\hline
7&0.00003 	&0.00012 	&0.00003 	&0.00004 	&0.00055 	&0.00020 &21325 \\
\hline
8&0.00003 	&0.00017 	&0.00006 	&0.00008 	&0.00170 	&0.00047 &49707 \\
\hline
9 &0.00004 	&0.00024 	&0.00008 	&0.00007 	&0.00155 	&0.00052 &109904\\
\hline
10 &0.00006 	&0.00049 	&0.00014 	&0.00012 	&0.00289 	&0.00077 &254730 \\
\hline
11&0.00009 	&0.00100 	&0.00027	&0.00014 	&0.00971 	&0.00213 &315316\\
\hline
12& 0.00010 	&0.00164 	&0.00030 	&0.00015 	&0.01746 	&0.00233 &374120 \\
\hline
13 & 0.00012 	&0.00071 	&0.00034 	&0.00018	&0.01273 	&0.00325 &411656  \\
\hline
14 & 0.00020 	&0.00359 	&0.00061 	&0.00025 	&0.01942 	&0.00375 &487357 \\
\hline
15 & 0.00035 	&0.00307 	&0.00073	&0.00030 	&0.02915 	&0.00428 &502045  \\
\hline
16& 0.00056 	&0.00391	&0.00096 	&0.00041 	&0.07559 	&0.00689 &580102  \\
\hline
17 & 0.00514 	&0.00730	&0.00175 	&0.00090 	&0.10188 	&0.01278 &572623 \\
\hline
18 & 0.00218 	&0.02287 	&0.00540 	&0.00168 	&0.57248 	&0.01797 &576476 \\
\hline
19 & 0.00400 	&0.05346	&0.00760	&0.00263 	&0.22828 	&0.02507 &547585 \\
\hline
20 & 0.00764 	&0.74596 	&0.01725 	&0.00498 	&0.00986 	&0.02535 &630036 \\
\hline
21 & 0.01503 	&0.03357 	&0.04729 	&0.00960 	&0.01915	&0.05746  &626900  \\
\hline
22& 0.02965 	&0.06603 	&0.050152 	&0.01805 	&0.03224 	&0.07098 &738381 \\
\hline
\end{tabular}
\end{table}

\begin{figure}[!h]
\centering
\includegraphics[width=0.4\textwidth,height=7cm]{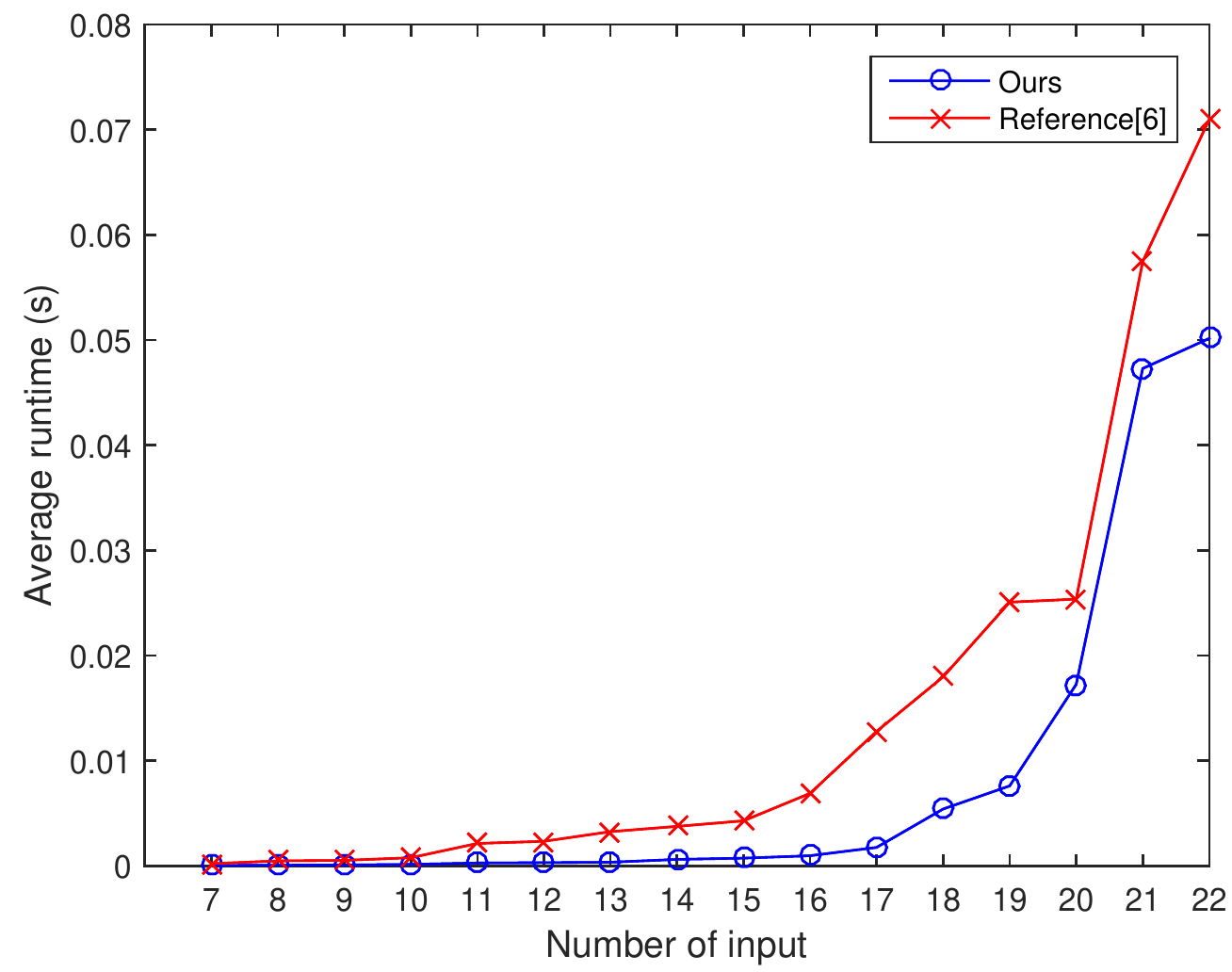}
\caption{Comparison of average runtimes on the first type of equivalent random circuits}
\label{fig.4}
\end{figure}

The matching speed presented in Table \uppercase\expandafter{\romannumeral2} is faster than that shown in Table \uppercase\expandafter{\romannumeral1} obviously.
This result occurs because the experiment on the first type of equivalent random circuits generates fewer multiple-mapping sets than that on the equivalent MCNC benchmark circuits. Greater numbers of multiple-mapping sets and larger cardinality will yield more branches; consequently, Procedure 2 must test whether $f(TX)=g(X)$ or $f(TX)=\overline{g(X)}$ many more times. The average runtime of our algorithm is 5.6 times faster than competitors when testing the equivalent circuits on the first type of equivalent random circuits.

Table \uppercase\expandafter{\romannumeral3} shows the results obtained for the second type of equivalent random circuits. Fig. 5 shows the average runtime of our algorithm compared to the average runtime of \cite{3} for the second type of equivalent random circuits.

\begin{table}[h]
\label{Tab3}
\caption{Boolean matching runtimes on the second type of equivalent random circuits}
\centering

\begin{tabular}{|p{0.2cm}<{\centering}|p{0.75cm}<{\centering}| p{0.8cm}<{\centering}|p{0.8cm}<{\centering}|p{0.8cm}<{\centering}| p{0.8cm}<{\centering}|p{0.8cm}<{\centering}|p{0.7cm}<{\centering}|}
\hline
\#I &\#MIN & \#MAX &\#AVG  &\#MIN of Ref\cite{3} & \#MAX of Ref\cite{3} & \#AVG of Ref\cite{3} & \#AVG nodes\\
\hline
7&0.00008 	&0.00079 	&0.00018 	&0.00025 	&0.00108 	&0.00063 &16811 \\
\hline
8 &0.00014 &0.00109 	&0.00024 	&0.00055 	&0.00214 	&0.00128 &39696\\	
\hline
9&0.00013 	&0.00256 	&0.00070 	&0.00105 	&0.00381 	&0.00281 &87675\\	
\hline
10&0.00071 	&0.00504 	&0.00118 	&0.00244 	&0.00824 	&0.00474 &196241\\
\hline
11 &0.00159 	&0.00957	&0.00243 	&0.00244 	&0.01278 	&0.00730 &423826\\
\hline
12 & 0.00059 	&0.01470 	&0.00715 	&0.01427 	&0.02630 	&0.02069  &512004\\
\hline
13 & 0.01326 	&0.02894 	&0.01312 	&0.02940 	&0.06153 	&0.04290  &514216\\	
\hline
14 & 0.03208 	&0.06498 	&0.02976 	&0.06475 	&0.11262 	&0.08946  &537731\\
\hline
15 & 0.07196 	&0.13799 	&0.06567 	&0.19312 	&0.27239 	&0.18574  &550951\\
\hline
16 & 0.15473 	&0.26420 	&0.15203 	&0.35803 	&0.54711 	&0.40130 &542536 \\
\hline
17 & 0.38728 	&0.73046 	&0.34597 	&0.79526 	&1.00914 	&0.84554 &542037\\
\hline
18& 0.74616 	&2.00846 	&0.79530 	&1.66938 	&2.10486 	&1.78664  &557623\\
\hline
19 & 2.13903 	&3.24150 	&1.97701 	&3.81065 	&4.14354 	&3.97135  &562577\\
\hline
20 & 5.40300 	&8.86237 	&4.73963 	&9.13982 	&16.64080 	&15.76470  &649762\\
\hline
21 & 10.28530 	&16.76480 	&10.53660 	&23.56350 	&28.96760 	&26.18972  &729953\\	
\hline
22& 35.15610 	&93.85770 	&38.0587 	&58.18920 	&99.24210 	&74.63841  &836984\\
\hline
\end{tabular}
\end{table}

\begin{figure}[!h]
\centering
\includegraphics[width=0.4\textwidth,height=7cm]{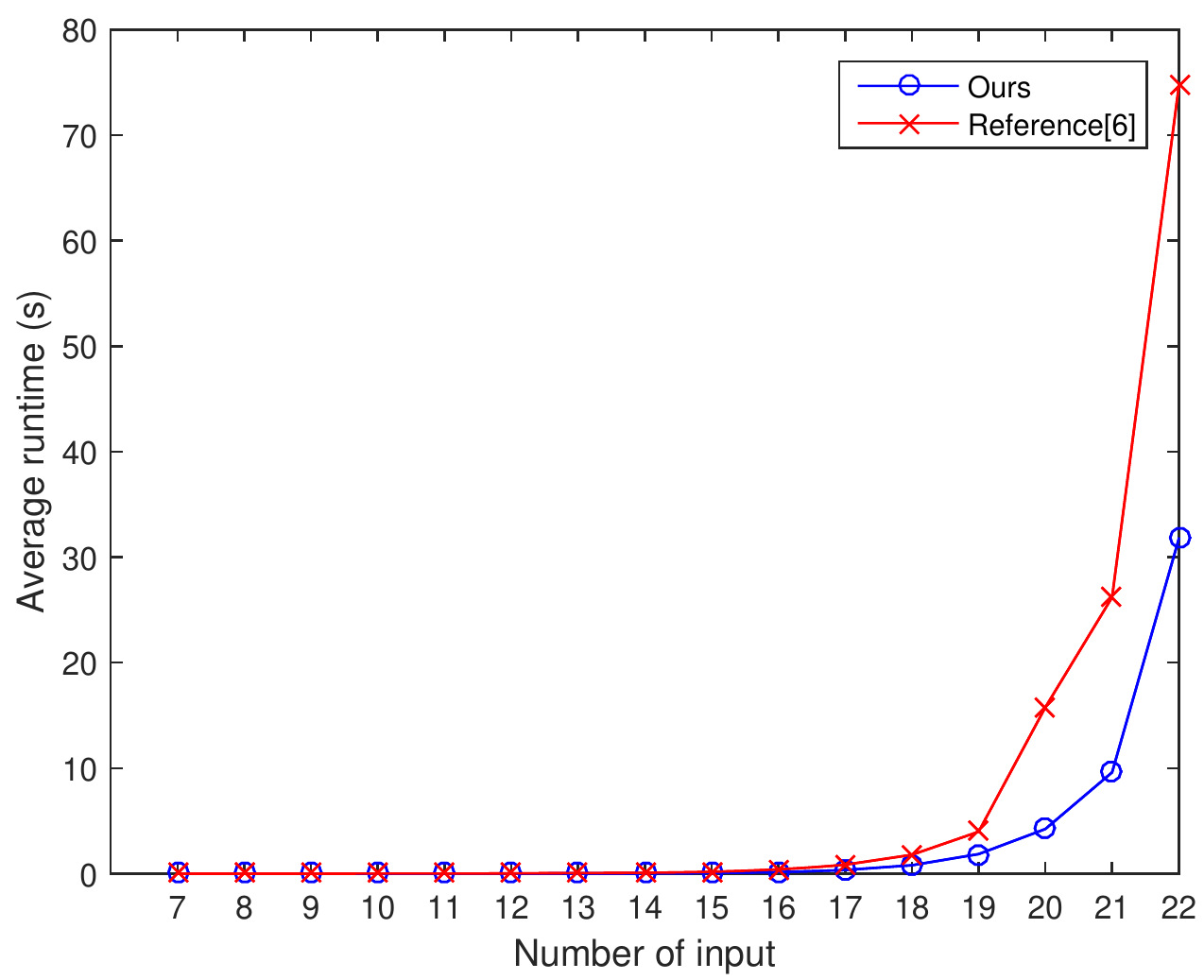}
\caption{Comparison of average runtimes on the second type of equivalent random circuits}
\label{fig.5}
\end{figure}

The number of minterms for the second type of random circuits is $2^{n-1}$. Our algorithm first matches $f$ and $g$ and then matches $f$ and $\overline{g}$ when $f$ and $g$ are not NP-equivalent. Compared with Tables \uppercase\expandafter{\romannumeral1} and \uppercase\expandafter{\romannumeral2}, the runtime presented in Table \uppercase\expandafter{\romannumeral3} is significantly longer than those in Tables \uppercase\expandafter{\romannumeral1} and \uppercase\expandafter{\romannumeral2}. This result occurs because transformation detection is more likely to be executed twice. The results of Table \uppercase\expandafter{\romannumeral3} show that the average runtime of our algorithm is 3.1 times faster than competitors when testing the second type of equivalent random circuits.

Tables \uppercase\expandafter{\romannumeral1}, \uppercase\expandafter{\romannumeral2} and \uppercase\expandafter{\romannumeral3} reveal that our algorithm is faster than that of \cite{3}. On average, our runtime is generally 4.2 times faster than that of \cite{3} when tested on equivalent functions. Nevertheless, when we tested several 17-input circuits from the MCNC benchmark, the matching speed of the algorithm of \cite{3} was revealed to be faster than that of ours. In the test of the Boolean functions of these 17-input circuits, there are multiple variables whose SS values were always the same, and whose $1^{st}$ signature values are the power of 2 and they are asymmetric variable.

Clearly, the space complexity of the worst circuit is equal to that of the exhaustive method, i.e., an exponential in $n$. Currently, no effective methods for solving the Boolean matching of the worst circuits exists. However, the execution speed of our algorithm is generally linear with respect to the number of inputs for general circuits. Overall, our algorithm is superior to that in \cite{3} regarding NPN Boolean matching for general circuits.

We also tested our algorithm on non-equivalent circuits. Table \uppercase\expandafter{\romannumeral4} shows the experimental results for the non-equivalent MCNC benchmark circuits. Fig. 6 shows the average runtime of our algorithm compared with the average runtime of \cite{3} for the non-equivalent MCNC benchmark circuits.

\begin{table}[h]
\label{Tab4}
\caption{Boolean matching runtimes on the non-equivalent MCNC benchmark circuits}
\centering

\begin{tabular}{|p{0.3cm}|p{0.8cm}| p{0.8cm}|p{0.8cm}|p{0.8cm}| p{0.8cm}|p{0.8cm}|}
\hline
\#I &\#MIN & \#MAX &\#AVG  &\#MIN of Ref\cite{3} & \#MAX of Ref\cite{3} & \#AVG of Ref\cite{3} \\
\hline
7  &0.00001 	&0.00024 	&0.00005 	&0.00004 	&0.03633 	&0.00104 \\
\hline
8  &0.00003 	&0.00024 	&0.00010 	&0.00028 	&0.00601 	&0.00128 \\
 \hline
9  &0.00003 	&0.00026    &0.00011 	&0.00006 	&0.00248 	&0.00145 \\
\hline
10 &0.00004 	&0.00046 	&0.00018 	&0.00026 	&0.00350 	&0.00172 \\
\hline
11 &0.00005 	&0.00033 	&0.00020 	&0.00030 	&0.00418 	&0.00260 \\
\hline
12 &0.00006 	&0.00053 	&0.00020 	&0.00030 	&0.02120 	&0.00208 \\
\hline
13 &0.00023 	&0.00102 	&0.00040 	&0.00111 	&0.10677 	&0.00234 \\
\hline
14 &0.00020 	&0.00253 	&0.00088 	&0.00135 	&0.04268 	&0.03071 \\
\hline
15 &0.00010 	&0.00313 	&0.00085 	&0.00076 	&0.11391 	&0.01109  \\
\hline
16 &0.00011 	&0.00232 	&0.00086 	&0.00080 	&0.00476 	&0.03865 \\
\hline
17 &0.00007 	&0.00257 	&0.00088 	&0.00071 	&1.05822 	&0.30897 \\
\hline
18 &0.00006 	&0.00479 	&0.00117	&0.00124 	&3.02963 	&0.37848  \\
\hline
19 &0.00012 	&0.01434 	&0.00273 	&0.00304 	&3.76650 	&1.53924  \\
\hline
20&0.00066 	    &0.00898 	&0.00279 	&0.19141 	&10.11520 	&2.29006 \\
\hline
21 &0.00091 	&0.00678 	&0.00339 	&0.33115 	&10.36450 	&3.46260 \\
\hline
22 &0.00113 	&0.02299 	&0.00697 	&9.15070 	&27.38330 	&12.84580 \\
\hline
\end{tabular}
\end{table}

\begin{figure}[!h]
\centering
\includegraphics[width=0.4\textwidth,height=7cm]{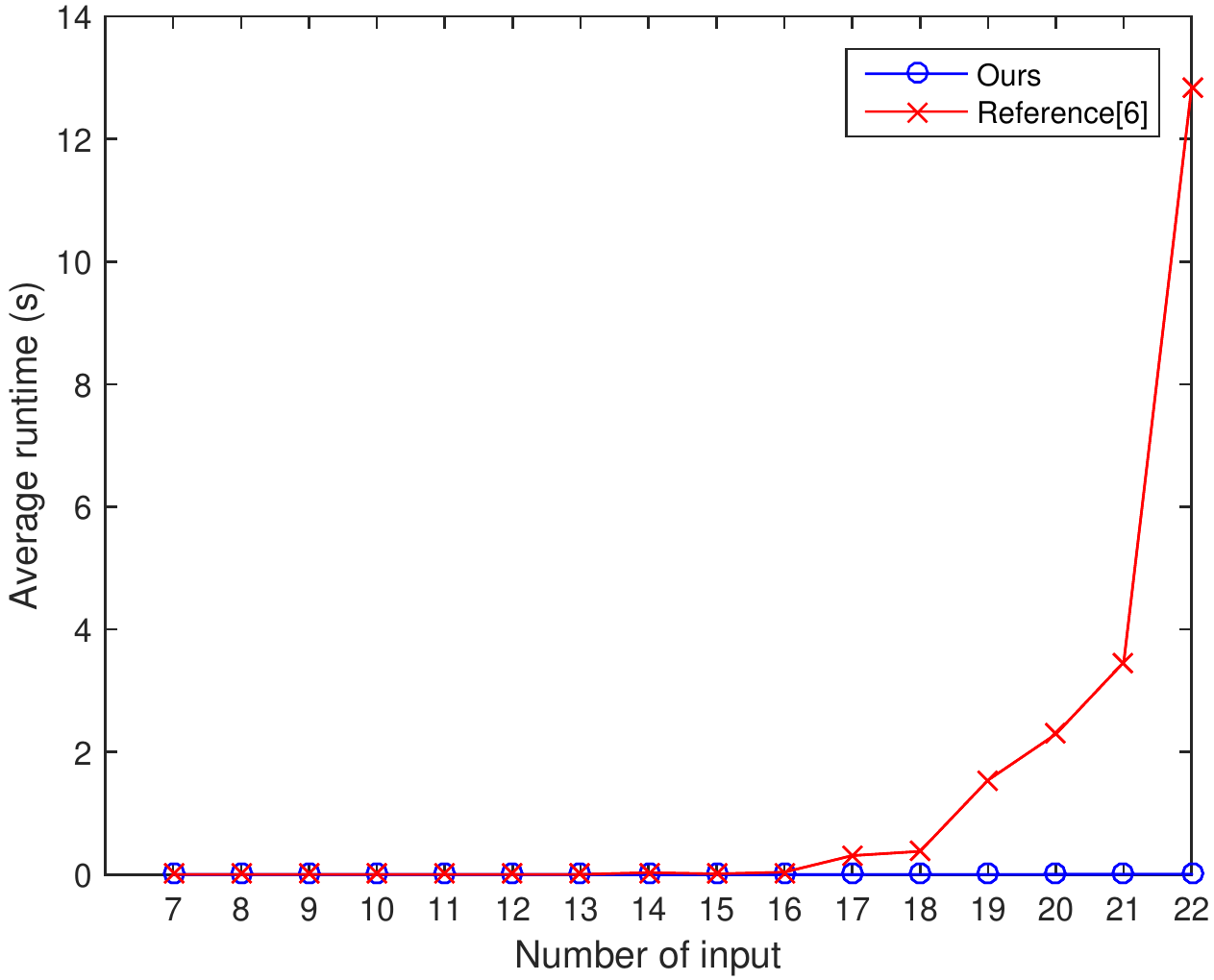}
\caption{Comparison of average runtimes on the non-equivalent MCNC benchmark circuits}
\label{fig.6}
\end{figure}

The results of Table \uppercase\expandafter{\romannumeral4} show the matching speed of our algorithm on non-equivalent MCNC benchmark circuits is far faster than that of the algorithm of \cite{3}. When the number of inputs is 22, the average runtime of our algorithm is 1,843 times faster than that of the algorithm from \cite{3}.

 Table \uppercase\expandafter{\romannumeral5} lists the experimental results for the non-equivalent random circuits, and Fig. 7 shows the average runtime of our algorithm compared with the average runtime of \cite{3} for the non-equivalent random circuits.

\begin{table}[h]
\label{Tab5}
\caption{Boolean matching runtimes on the non-equivalent random circuits}
\centering

\begin{tabular}{|p{0.3cm}|p{0.8cm}| p{0.8cm}|p{0.8cm}|p{0.8cm}| p{0.8cm}|p{0.8cm}|}
\hline
\#I &\#MIN & \#MAX &\#AVG  &\#MIN of Ref\cite{3} & \#MAX of Ref\cite{3} & \#AVG of Ref\cite{3} \\
\hline
7  &0.00001 	&0.00010 	&0.00004 	&0.00005 	&0.00045 	&0.00010 \\
\hline
8  &0.00002 	&0.00016 	&0.00005 	&0.00007 	&0.00163 	&0.00019 \\
 \hline
9  &0.00002 	&0.00018    &0.00005 	&0.00007 	&0.00122 	&0.00035 \\
\hline
10 &0.00002 	&0.00019 	&0.00006 	&0.00014 	&0.00231 	&0.00066 \\
\hline
11 &0.00004 	&0.00024 	&0.00007 	&0.00014 	&0.00832	&0.00193 \\
\hline
12 &0.00004 	&0.00038 	&0.00011 	&0.00013 	&0.02332 	&0.00235 \\
\hline
13 &0.00003 	&0.00073 	&0.00011 	&0.00014 	&0.01227 	&0.00253 \\
\hline
14 &0.00005 	&0.00032 	&0.00012 	&0.00024 	&0.01643 	&0.00325 \\
\hline
15 &0.00007 	&0.00073 	&0.00019 	&0.00029 	&0.02793 	&0.00375  \\
\hline
16 &0.00004 	&0.00102 	&0.00021 	&0.00041 	&0.09054 	&0.00871 \\
\hline
17 &0.00004 	&0.00104 	&0.00026 	&0.00097 	&0.09840 	&0.01260 \\
\hline
18 &0.00013 	&0.00268 	&0.00032	&0.00198 	&0.40210 	&0.01305  \\
\hline
19 &0.00011 	&0.00417 	&0.00053 	&0.00273 	&0.42959 	&0.01907  \\
\hline
20&0.00017 	    &0.00057 	&0.00054 	&0.00497 	&0.16032 	&0.02594 \\
\hline
21 &0.00018 	&0.03548 	&0.00174 	&0.00730 	&0.01732 	&0.03149 \\
\hline
22 &0.00018 	&0.03426 	&0.00195 	&0.01855 	&0.02014 	&0.05781 \\
\hline
\end{tabular}
\end{table}

\begin{figure}[!h]
\centering
\includegraphics[width=0.4\textwidth,height=7cm]{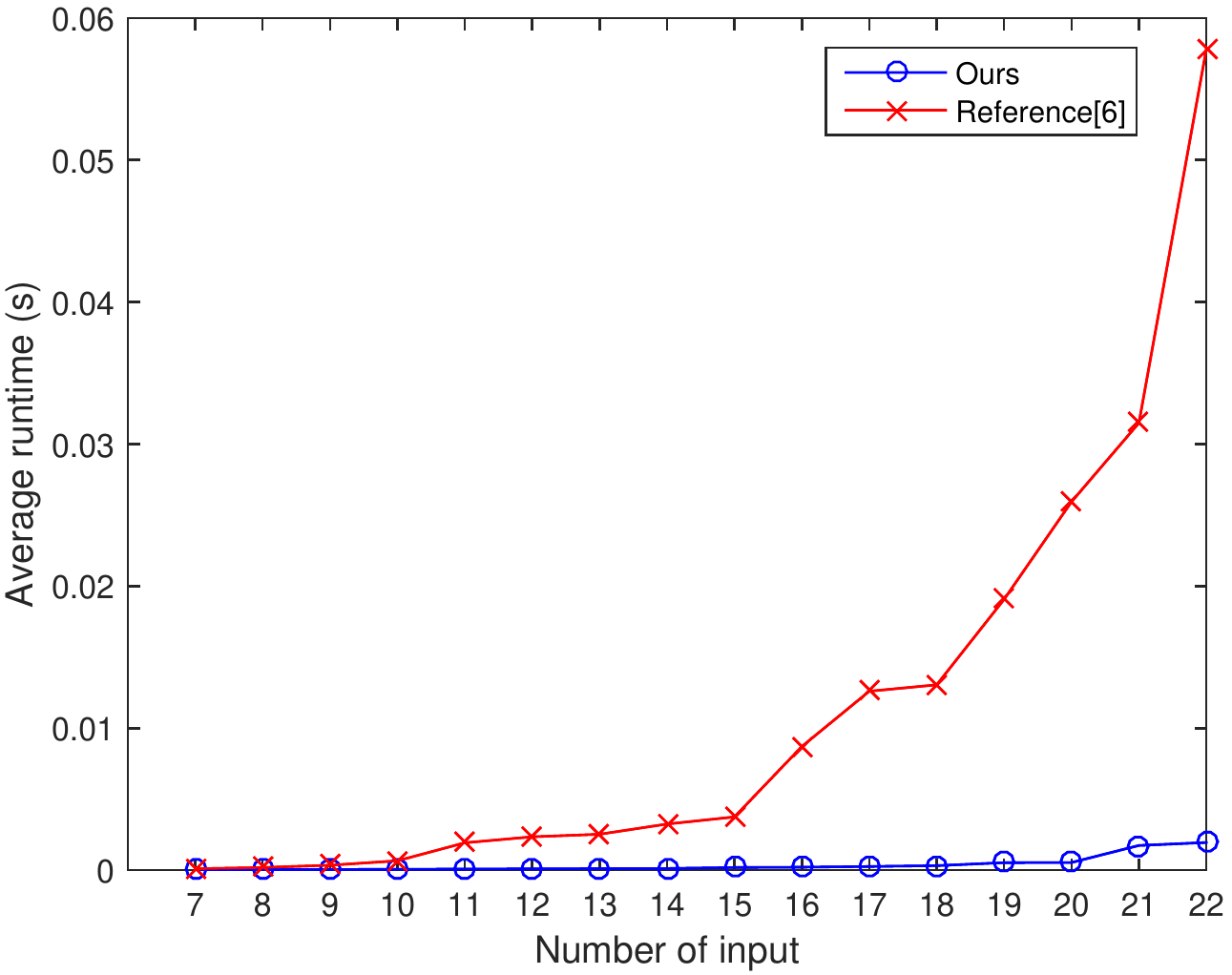}
\caption{Comparison of average runtimes on the non-equivalent random circuits}
\label{fig.7}
\end{figure}

As shown in Figs. 6--7 and Tables \uppercase\expandafter{\romannumeral4} and \uppercase\expandafter{\romannumeral5}, the average runtime of our algorithm is averagely 172 times faster than that of \cite{3}. As the number of inputs increases, the average runtime of our algorithm becomes exceptionally shorter than that of \cite{3}.

In the process of non-equivalent circuit matching, three possible cases exist: (1) Matching terminates after the first equivalence judgement of two SS vectors occurs. (2) Our algorithm does not find any possible transformations. (3) Our algorithm finds some possible transformations, but none of these transformations satisfy $f(TX)=g(X)$ or $f(TX)=\overline{g(X)}$. In our testing, almost all non-equivalent circuit functions belong to (1), and a handful of non-equivalent circuits belong to (2). Our algorithm spends considerable time generating circuit functions that belong to (3). We generated a small number of these circuit functions and tested them. The runtime of our algorithm for non-equivalent circuit matching is shorter than that for equivalent circuit matching because the latter involves very few possible transformations. In contrast, the algorithm of \cite{3} must compute the canonical form twice regardless. Even when the canonical form of circuit functions is stored in a cell library in advance, the algorithm of \cite{3} is slower than ours because it still must compute one canonical form. Therefore, our algorithm has an obvious advantage over the algorithm of \cite{3} when matching non-equivalent Boolean functions.

According to the non-equivalent matching results, the performance of our algorithm is more robust than that of \cite{3}.
Specifically, as shown in Table \uppercase\expandafter{\romannumeral4}, when the number of inputs is 22, the average runtime of our algorithm is approximately 140 times greater than when the number of inputs is 7, while the average runtime of the algorithm of \cite{3} is approximately 12351 times greater than that when the number of inputs is 7.

The experimental results illustrate the effectiveness of the proposed algorithm and demonstrate that it can be applied to Boolean matching for large-scale circuits.

\section{Conclusions}

This paper proposes an efficient NPN Boolean matching algorithm based on a new SS vector and the Shannon expansion theorem. The proposed algorithm prunes the search space and significantly reduces the time complexity.

 Compared with the algorithm of \cite{3}, our algorithm is 4.2 times faster for general equivalent circuits and 172 times faster, on average, for non-equivalent circuits. Our algorithm is exceptionally faster at non-equivalent matching and is more robust compared with the algorithm of \cite{3}. The algorithm can be used for both large-scale circuit technology mapping and cell-library binding. In future work, we will apply our algorithm to Boolean matching with don't care sets and to multiple-output Boolean function matching.

\ifCLASSOPTIONcaptionsoff
  \newpage
\fi



%

\section*{Acknowledgment}

We would like to thank the National Natural Science Foundation of China (Grant No. 61572109) for the technology support.

\bibliographystyle{./IEEEtran}

\bibliography{./IEEEabrv,./paper}

\end{document}